\RequirePackage{tikz}
\RequirePackage{fix-cm}
\documentclass[review]{elsarticle}

\usepackage{blindtext}
\usepackage[hidelinks]{hyperref}
\usepackage[absolute,overlay]{textpos}
\usepackage[margin=1.44in]{geometry}
\usepackage{hypernat}

\usepackage{amssymb}
\usepackage{amsmath}
\usepackage{bm}
\usepackage{shellesc}
\usepackage{subcaption}
\usetikzlibrary{arrows,shapes,decorations.pathreplacing,calligraphy}
\usepackage{booktabs}
\usepackage{adjustbox}
\usepackage{siunitx}
\usepackage{upgreek} 
\usepackage[utf8]{inputenc}
\usepackage[T1]{fontenc}
\usepackage{lmodern}
\usepackage{etoolbox}
\usepackage{algorithm}
\usepackage{algpseudocode}
\usepackage{siunitx}
\usepackage{physics}
\usepackage{mathdots}
\usepackage{yhmath}
\usepackage{booktabs}
\usepackage{gensymb}
\usepackage{wasysym}
\usepackage{multirow}
\usepackage{pgfplotstable}
\usepgfplotslibrary{polar} 
\usepackage{color}
\usepackage{multirow}
\usepackage{array}
\usepgfplotslibrary{external}
\usetikzlibrary{fadings}
\usetikzlibrary{patterns}
\usetikzlibrary{shadows.blur}
\usetikzlibrary{arrows}
\usetikzlibrary{arrows.meta}
\usetikzlibrary{shapes}
\usepackage{algorithmicx}
\usetikzlibrary{shapes.geometric, positioning, fit, backgrounds}

\usepackage{tikz}
\usetikzlibrary{external}
\tikzset{external/system call={lualatex -shell-escape -halt-on-error 
		-interaction=batchmode -jobname "\image" "\texsource" &&
		pdftops -eps "\image.pdf" "\image.eps"}}
\usepackage{pgfplots}
\pgfplotsset{compat=1.17}

\usepackage[]{caption}[2008/08/24]
\definecolor{cbgreen}{RGB}{34,139,34}

\definecolor{tealfill}  {RGB}{225,245,238}
\definecolor{tealbd}    {RGB}{15, 110, 86}
\definecolor{purplefill}{RGB}{238,237,254}
\definecolor{purplebd}  {RGB}{83, 74, 183}
\definecolor{coralfill} {RGB}{250,236,231}
\definecolor{coralbd}   {RGB}{153, 60, 29}
\definecolor{bluefill}  {RGB}{230,241,251}
\definecolor{bluebd}    {RGB}{24, 95, 165}
\definecolor{amberfill} {RGB}{250,238,218}
\definecolor{amberbd}   {RGB}{133, 79, 11}
\definecolor{grayfill}  {RGB}{241,239,232}
\definecolor{graybd}    {RGB}{95, 94, 90}
\definecolor{phasebg}   {RGB}{248,248,248}
\definecolor{phaseborder}{RGB}{180,178,170}

\tikzset{
	base/.style={
		rectangle, rounded corners=6pt,
		text width=3.8cm, align=center,
		minimum height=1.3cm,
		font=\small,
		inner sep=6pt,
		line width=0.4pt,
	},
	tealbox/.style  ={base, fill=tealfill,   draw=tealbd},
	purplebox/.style={base, fill=purplefill, draw=purplebd,
		text width=8cm, minimum height=1.5cm},
	coralbox/.style ={base, fill=coralfill,  draw=coralbd},
	bluebox/.style  ={base, fill=bluefill,   draw=bluebd,
		text width=7.5cm, minimum height=1.3cm},
	amberbox/.style ={base, fill=amberfill,  draw=amberbd,
		text width=4.5cm, minimum height=1.3cm},
	graybox/.style  ={base, fill=grayfill,   draw=graybd},
	phaselabel/.style={font=\small\itshape, text=phaseborder},
	arr/.style={-{Stealth[length=5pt,width=4pt]},
		line width=0.6pt, color=gray!70},
	dasharr/.style={arr, dashed},
	thinline/.style={line width=0.6pt, color=gray!70},
}

\journal{Composites Part B: Engineering}

\begin{document}

\begin{frontmatter}
		
\title{\fontsize{12}{16}\selectfont Data-Driven Structural Health Monitoring of Short Carbon Fiber-Reinforced Polymer Composites via Multiphysics Phase-Field Simulation}

%
\author[1,2]{Behrouz Arash\corref{cor1}}
\ead{behrouza@oslomet.no}
\author[1]{Shadab Zakavati}
\author[3]{Quan Wang}
\ead{wangquan@stu.edu.cn}
\author[4]{Timon Rabczuk}
\ead{timon.rabczuk@uni-weimar.de}
\address[1]{Department of Mechanical, Electrical and Chemical Engineering, Oslo Metropolitan University, Pilestredet 35, 0166 Oslo, Norway}
\address[2]{Green Energy Lab, Department of Mechanical, Electrical and Chemical Engineering, OsloMet - Oslo Metropolitan University, Oslo, Norway}
\address[3]{College of Engineering, Shantou University, Shantou, Guangdong 515063, China}
\address[4]{Institute of Structural Mechanics, Bauhaus-Universit{\"a}t Weimar, Marienstra{\ss}e 15, 99423 Weimar, Germany}

\cortext[cor1]{Corresponding author}



\begin{abstract}
Short carbon fiber-reinforced polymer (SCFRP) composites exploit the intrinsic conductivity of the carbon fiber network for self-sensing, yet no predictive model couples their anisotropic, rate-dependent fracture to piezoresistive damage identification. This work presents a finite-deformation multiphysics phase-field framework coupling a viscoelastic-viscoplastic constitutive model, an anisotropic crack resistance formulation, and a piezoresistive conductivity model. The three sub-problems are unified through the second-order fiber orientation tensor, which simultaneously defines fiber family directions, crack resistance anisotropy, and principal conduction paths of the carbon fiber network. A damage-coupled conductivity tensor captures both strain-driven geometric-kinematic resistance changes and irreversible network severance driven by the phase-field variable. The framework is coupled to an eight-electrode electrical impedance tomography configuration, and the normalized inter-electrode conductance ratios serve as inputs to a feedforward artificial neural network that infers normalized crack length and mechanical compliance without mechanical sensing. 
The network achieves $R^2 = 0.99$ on held-out configurations, confirming generalization across the microstructure space. The framework establishes a physics-based, computationally efficient route for real-time structural health monitoring and inverse damage assessment in SCFRP composites.
\end{abstract}

\end{frontmatter}


\section{Introduction}
\label{sec:sec1}

Short carbon fiber-reinforced polymer (SCFRP) composites have become indispensable
in aerospace, automotive, and renewable energy applications owing to their high
specific stiffness and strength, tailorable anisotropy, and the intrinsic electrical
conductivity of the carbon fiber (CF)
network~\cite{holbery2006natural,soutis2005carbon,arash2019viscoelastic2}.
This electrical conductivity gives SCFRP composites a dual functionality: they serve
simultaneously as load-bearing structural members and as self-sensing elements
capable of detecting internal damage through changes in their electrical
resistance~\cite{chung2001continuous,wang2006self}.
Realizing this self-sensing potential in service, however, requires predictive models
that couple the anisotropic, rate-dependent mechanical behavior and fracture of the
composite to its evolving electrical response, a coupling that remains
largely absent from the literature.

The mechanical response of SCFRP composites exhibits pronounced anisotropy governed
by fiber orientation distributions, which controls not only elastic properties but
also inelastic deformation mechanisms and fracture
behavior~\cite{gusev2017finite,wang2019numerical,arash2019viscoelastic2}.
Polymer matrices additionally exhibit rate-dependent behavior through coupled
viscoelastic and viscoplastic
mechanisms~\cite{poulain2014finite,arash2021finite,arash2025phase}, and hygrothermal
conditions further degrade material performance through moisture absorption and
thermal
cycling~\cite{zhou2007experimental,bahtiri2023machine,arash2025phase}.
The interplay between mechanical anisotropy, rate-dependent inelasticity,
environmental degradation, and damage evolution presents formidable challenges for
predictive modeling frameworks.

Phase-field modeling has emerged as a variational framework for fracture that
overcomes the limitations of cohesive zone
models~\cite{ortiz1999finite,li2005use}, element deletion
techniques~\cite{lapczyk2007progressive}, and extended finite element
methods~\cite{pike2015xfem,kastner2016xfem} by representing cracks through a
continuous damage field, eliminating explicit crack tracking and providing
mesh-independent
solutions~\cite{francfort1998revisiting,bourdin2000numerical,miehe2010phase,%
	ambati2015review,wu2020phase}.
The framework has been extended to
brittle~\cite{miehe2010thermodynamically,borden2012phase},
ductile~\cite{ambati2015phase,miehe2016phase,dean2026hybrid}, cohesive
fracture~\cite{verhoosel2013phase}, anisotropic fracture
energy~\cite{li2015phase,teichtmeister2017phase,clayton2015phase}, dynamic
effects~\cite{hofacker2013phase}, and multi-field
coupling~\cite{wilson2013phase,miehe2015phase,arash2025phase}.

Anisotropic phase-field models for fiber-reinforced composites introduce
directionally dependent fracture energy that reflects preferential crack paths
imposed by fiber
orientation~\cite{bleyer2018phase,gultekin2016phase,dean2020multi,lv2026anisotropic}.
Structural tensor-based formulations capture orientation-dependent fracture
resistance~\cite{li2015phase,teichtmeister2017phase}, and multi-phase-field
frameworks resolve simultaneously intralaminar cracking and interlaminar delamination
in laminated
composites~\cite{dean2020multi,dean2020phase,dean2026hybrid}.
Phase-field models incorporating hygro-mechanical coupling have also
emerged~\cite{au2023hygroscopic,arash2025phase}, though most existing anisotropic
formulations consider elastic or small-strain elasto-plastic
behavior~\cite{duda2015phase,ulloa2019phase}, with limited attention to the coupled
viscoelastic-viscoplastic response of polymer matrices at finite
deformation~\cite{dammass2023phase,kumar2022nonlinear}.
For polymer-based materials, phase-field frameworks incorporating viscoelastic
behavior have been
developed~\cite{yin2022viscoelastic,shen2019fracture,loew2019rate,brighenti2021phase},
and Arash and coworkers~\cite{arash2021finite,arash2023effect,arash2026phase} 
extended these to finite-deformation viscoelastic-viscoplastic fracture with
hygrothermal coupling for short fiber-reinforced polymer composites.
The combination of viscoelastic and viscoplastic mechanisms within phase-field
fracture frameworks has not, however, been coupled to the electrical self-sensing
response of the carbon fiber network, and no existing framework integrates
this mechanical-fracture model with inverse damage identification for structural
health monitoring.

The piezoresistive self-sensing capability of SCFRPs has been studied
experimentally for several
decades~\cite{chung2001continuous,abry1999situ,wang2006self}.
Two physically distinct mechanisms contribute to resistance changes under mechanical
loading: a geometric-kinematic mechanism, in which deformation alters effective
conduction path lengths and fiber contact geometry; and a damage-driven mechanism,
in which matrix cracking and fiber-matrix debonding irreversibly sever conductive
pathways~\cite{abry1999situ,todoroki2004electrical}.
Despite this rich experimental foundation, predictive computational models coupling
these mechanisms to a continuum fracture framework remain scarce.
Existing electrical models typically assume linear piezoresistivity with prescribed
damage~\cite{gallo2015electrical} and do not account for the anisotropic,
orientation-dependent conductivity tensor or its evolution with the phase-field
damage variable.
Furthermore, no study has combined piezoresistive phase-field simulation with a
multi-electrode measurement setup and machine learning inversion for real-time
structural health monitoring of SCFRP composites.

Motivated by these considerations, the present work develops an integrated
multiphysics phase-field framework for modeling anisotropic
viscoelastic-viscoplastic fracture and damage-induced piezoresistive response in
SCFRP composites at finite deformation, and combines it with a data-driven inverse
framework for structural health monitoring.
The principal contributions are as follows.
First, a finite-deformation viscoelastic-viscoplastic constitutive model is
formulated through multiplicative decomposition of the deformation gradient,
extending the framework of
Arash et al.~\cite{arash2021finite,arash2023effect,arash2026phase} to incorporate
piezoresistive self-sensing in anisotropic short fiber composites.
Second, an anisotropic phase-field fracture formulation employing the second-order
fiber orientation tensor captures orientation-dependent fracture energy for arbitrary
fiber architectures.
Third, a damage-coupled piezoresistive conductivity tensor captures both
geometric-kinematic resistance changes through axial and transverse gauge factors
and irreversible damage-driven degradation consistent with the phase-field model.
The fiber orientation tensor serves as the common thread: it simultaneously defines
fiber family directions, anisotropic crack resistance, and principal conduction paths
of the carbon fiber network.
Fourth, the framework is combined with an eight-electrode electrical impedance tomography (EIT) setup and a feedforward artificial neural network (ANN) for real-time inverse
damage identification from conductance measurements alone.

The remainder of this paper is organized as follows.
Section~\ref{sec:const} presents the viscoelastic-viscoplastic constitutive model,
detailing the kinematics, stress decomposition, and fiber family representation.
Section~\ref{sec:phase} introduces the anisotropic phase-field fracture formulation.
Section~\ref{sec:elec} develops the piezoresistive model and electric field
formulation.
Section~\ref{sec:shm_inverse} presents the EIT configuration and the ANN-based inverse
framework for structural health monitoring (SHM).
Section~\ref{sec:results} presents numerical investigations examining the effects of
fiber architecture, orientation, content, and temperature on the coupled
electromechanical fracture response and SHM performance.
Section~\ref{sec:summary} summarizes the key findings and outlines future research
directions.

\section{Constitutive model for short carbon fiber-reinforced epoxy composites}
\label{sec:const}

\begin{figure}[H]
	\centering
	\begin{subfigure}[b]{0.49\textwidth}
		\centering
		\includegraphics[scale=0.85]{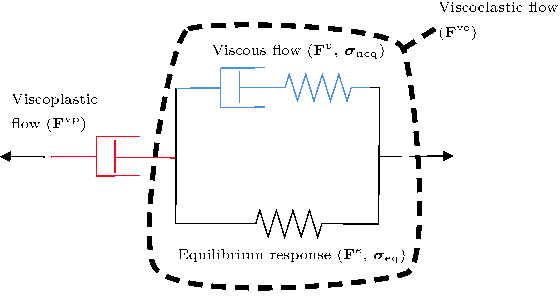}
		\caption{}
		\label{fig:rheological_model_a}
	\end{subfigure}%
	\hfill
	\begin{subfigure}[b]{0.49\textwidth}
		\centering
		\includegraphics[scale=0.85]{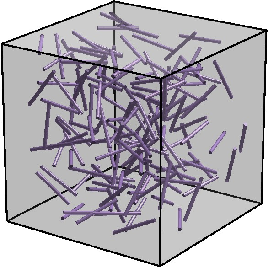}
		\caption{}
		\label{fig:rheological_model_b}
	\end{subfigure}
	\caption{(a) One-dimensional schematic of the viscoelastic-viscoplastic
		constitutive model for SCFRP composites. (b) Representative volume element
		showing the carbon fiber network embedded in the epoxy matrix.}
	\label{fig:rheological_model}
\end{figure}

A finite-deformation viscoelastic-viscoplastic constitutive model is developed for short carbon fiber-reinforced polymer (SCFRP) composites. The model combines the rate-dependent inelastic response of the epoxy matrix with the anisotropic stiffening contributed by multiple carbon fiber families. A schematic of the rheological model and the representative volume element are shown in Fig.~\ref{fig:rheological_model}. The deformation gradient $\mathbf{F}$, the viscous internal variable $\bar{\mathbf{F}}^v$, and the viscoplastic internal variable $\bar{\mathbf{F}}^{\text{vp}}$ collectively define the state of the material at each material point. As detailed below, the orientation of the carbon fiber families is encoded in a second-order orientation tensor, which serves simultaneously as input to the mechanical constitutive law and to the piezoresistive conductivity model introduced in Section~\ref{sec:elec}.

\subsection{Kinematics}

The total deformation gradient $\mathbf{F}$ is multiplicatively split into
volumetric and isochoric (deviatoric) parts as
\begin{equation}
	\mathbf{F} = J^{1/3}\bar{\mathbf{F}},
	\label{eq:F_decomp}
\end{equation}
where $J = \det[\mathbf{F}]$ is the volumetric Jacobian and $\bar{\mathbf{F}}$
is the distortional deformation gradient satisfying $\det[\bar{\mathbf{F}}] = 1$.
The Jacobian is additively split to account for mechanical compressibility
and thermal dilatation as
\begin{equation}
	J = J_m J_\theta,
	\label{eq:J_decomp}
\end{equation}
where $J_m$ is the purely mechanical volumetric change and
\begin{equation}
	J_\theta = 1 + \alpha_\theta (\theta - \theta_0)
	\label{eq:J_theta}
\end{equation}
captures isotropic thermal expansion, with $\alpha_\theta$ the thermal
expansion coefficient, $\theta$ the absolute temperature, and
$\theta_0 = 296$~K the reference temperature. In the present study,
uniform ambient temperature is assumed throughout the specimen,
so $J_\theta$ enters the formulation as a fixed scalar at the start
of each simulation.

The distortional deformation gradient is further decomposed multiplicatively
into viscoelastic and viscoplastic contributions following the framework
of~\cite{govindjee1997presentation}:
\begin{equation}
	\bar{\mathbf{F}} = \bar{\mathbf{F}}^{\text{ve}}\bar{\mathbf{F}}^{\text{vp}},
	\label{eq:Fbar_decomp}
\end{equation}
where $\bar{\mathbf{F}}^{\text{ve}}$ and $\bar{\mathbf{F}}^{\text{vp}}$ denote
the viscoelastic and viscoplastic parts of the distortional deformation
gradient, respectively.
The viscoelastic component is itself split into elastic and viscous parts:
\begin{equation}
	\bar{\mathbf{F}}^{\text{ve}} = \bar{\mathbf{F}}^e\bar{\mathbf{F}}^v.
	\label{eq:Fve_decomp}
\end{equation}

The left Cauchy--Green deformation tensors associated with each intermediate
configuration are
\begin{align}
	\bar{\mathbf{B}} &= \bar{\mathbf{F}}\bar{\mathbf{F}}^T, \label{eq:Bbar}\\
	\bar{\mathbf{B}}^e &= \bar{\mathbf{F}}^e(\bar{\mathbf{F}}^e)^T, \label{eq:Be}\\
	\bar{\mathbf{B}}^{\text{ve}} &= \bar{\mathbf{F}}^{\text{ve}}(\bar{\mathbf{F}}^{\text{ve}})^T. \label{eq:Bve}
\end{align}

The velocity gradient of the viscoelastic network,
$\bar{\mathbf{L}}^{\text{ve}} = \dot{\bar{\mathbf{F}}}^{\text{ve}}(\bar{\mathbf{F}}^{\text{ve}})^{-1}$,
is decomposed into elastic and viscous contributions pushed forward to the
current configuration:
\begin{equation}
	\bar{\mathbf{L}}^{\text{ve}} = \bar{\mathbf{L}}^e
	+ \bar{\mathbf{F}}^e\bar{\mathbf{L}}^v(\bar{\mathbf{F}}^e)^{-1}
	= \bar{\mathbf{L}}^e + \tilde{\mathbf{L}}^v,
	\label{eq:Lve_decomp}
\end{equation}
where the viscous velocity gradient in the intermediate configuration is
\begin{equation}
	\tilde{\mathbf{L}}^v = \dot{\bar{\mathbf{F}}}^v(\bar{\mathbf{F}}^v)^{-1}
	= \tilde{\mathbf{D}}^v + \tilde{\mathbf{W}}^v.
	\label{eq:Lv_decomp}
\end{equation}
Here a tilde denotes quantities referred to the intermediate configuration,
$\tilde{\mathbf{D}}^v$ is the symmetric viscous stretching tensor, and
$\tilde{\mathbf{W}}^v$ is the skew-symmetric viscous spin tensor.
Following the standard convention~\cite{qi2005stress}, the intermediate
configuration is rendered unique by setting $\tilde{\mathbf{W}}^v = \mathbf{0}$.

The viscous stretching tensor is governed by the Argon flow rule~\cite{boyce1988large}:
\begin{equation}
	\tilde{\mathbf{D}}^v = \frac{\dot{\varepsilon}^v}{\tau_{\text{neq}}}
	\operatorname{dev}\!\left[\boldsymbol{\sigma}'_{\text{neq}}\right],
	\label{eq:Dv}
\end{equation}
where $\tau_{\text{neq}} = \|\operatorname{dev}[\boldsymbol{\sigma}_{\text{neq}}]\|_F$
is the Frobenius norm of the non-equilibrium stress deviator,
$\dot{\varepsilon}^v$ is the scalar viscous strain rate, and
$\boldsymbol{\sigma}'_{\text{neq}} = \mathbf{R}_e^T\boldsymbol{\sigma}_{\text{neq}}\mathbf{R}_e$
is the non-equilibrium Cauchy stress rotated to the stress-free intermediate
configuration via the elastic rotation $\mathbf{R}_e$ from the polar
decomposition $\bar{\mathbf{F}}^e = \mathbf{R}_e\mathbf{U}_e$.
The viscous strain rate takes the Arrhenius-type Argon form~\cite{arruda1993three}:
\begin{equation}
	\dot{\varepsilon}^v = \dot{\varepsilon}_0
	\exp\!\left[\frac{\Delta H}{k_b T}
	\left(\!\left(\frac{\tau_{\text{neq}}}{\tau_0}\right)^{\!m} - 1\right)\right],
	\label{eq:eps_dot_v}
\end{equation}
where $k_b$ is the Boltzmann constant, $\dot{\varepsilon}_0$ is the
pre-exponential strain rate, $\Delta H$ is the activation energy for viscous
flow, $\tau_0$ is the athermal shear resistance, and $m$ is a material
exponent~\cite{bahtiri2023machine}. From Eqs.~\eqref{eq:Lv_decomp}
and~\eqref{eq:Dv}, the evolution equation for the viscous deformation gradient
follows as
\begin{equation}
	\dot{\bar{\mathbf{F}}}^v
	= (\bar{\mathbf{F}}^e)^{-1}
	\frac{\dot{\varepsilon}^v}{\tau_{\text{neq}}}
	\operatorname{dev}\!\left[\boldsymbol{\sigma}'_{\text{neq}}\right]
	\bar{\mathbf{F}}^{\text{ve}}.
	\label{eq:Fv_evol}
\end{equation}

The velocity gradient of the total network,
$\bar{\mathbf{L}} = \dot{\bar{\mathbf{F}}}(\bar{\mathbf{F}})^{-1}$,
is likewise split into viscoelastic and viscoplastic parts:
\begin{equation}
	\bar{\mathbf{L}} = \bar{\mathbf{L}}^{\text{ve}}
	+ \bar{\mathbf{F}}^{\text{ve}}\bar{\mathbf{L}}^{\text{vp}}(\bar{\mathbf{F}}^{\text{ve}})^{-1}
	= \bar{\mathbf{L}}^{\text{ve}} + \tilde{\mathbf{L}}^{\text{vp}}.
	\label{eq:L_decomp}
\end{equation}
The viscoplastic velocity gradient is decomposed as
\begin{equation}
	\tilde{\mathbf{L}}^{\text{vp}}
	= \dot{\bar{\mathbf{F}}}^{\text{vp}}(\bar{\mathbf{F}}^{\text{vp}})^{-1}
	= \tilde{\mathbf{D}}^{\text{vp}} + \tilde{\mathbf{W}}^{\text{vp}},
	\label{eq:Lvp_decomp}
\end{equation}
with $\tilde{\mathbf{W}}^{\text{vp}} = \mathbf{0}$ imposed for uniqueness.
The viscoplastic stretching tensor is driven by the total deviatoric Cauchy
stress:
\begin{equation}
	\tilde{\mathbf{D}}^{\text{vp}}
	= \frac{\dot{\varepsilon}^{\text{vp}}}{\tau_{\text{tot}}}
	\operatorname{dev}[\boldsymbol{\sigma}],
	\quad
	\tau_{\text{tot}} = \|\operatorname{dev}[\boldsymbol{\sigma}]\|_F.
	\label{eq:Dvp}
\end{equation}
The viscoplastic strain rate is described by a threshold-based phenomenological
model:
\begin{equation}
	\dot{\varepsilon}^{\text{vp}} =
	\begin{cases}
		0 & \text{if } \tau_{\text{tot}} < \sigma_0, \\
		a(\varepsilon - \varepsilon_0)^b \dot{\varepsilon}
		& \text{if } \tau_{\text{tot}} \geq \sigma_0,
	\end{cases}
	\label{eq:eps_dot_vp}
\end{equation}
where $a$, $b$, and $\sigma_0$ are material constants, $\varepsilon_0$ marks
the onset of viscoplastic flow, $\varepsilon = \|\mathbf{E}\|_F$ is the
Frobenius norm of the Green--Lagrange strain tensor
\begin{equation}
	\mathbf{E} = \frac{1}{2}(\mathbf{F}^T\mathbf{F} - \mathbf{I}),
	\label{eq:Green_strain}
\end{equation}
and $\dot{\varepsilon}$ is the scalar strain rate. The corresponding
evolution of the viscoplastic deformation gradient is
\begin{equation}
	\dot{\bar{\mathbf{F}}}^{\text{vp}}
	= (\bar{\mathbf{F}}^{\text{ve}})^{-1}
	\frac{\dot{\varepsilon}^{\text{vp}}}{\tau_{\text{tot}}}
	\operatorname{dev}[\boldsymbol{\sigma}]\,\bar{\mathbf{F}}.
	\label{eq:Fvp_evol}
\end{equation}

\subsection{Implicit time integration}
\label{sec:time_integration}

The evolution Eqs.~\eqref{eq:Fv_evol} and~\eqref{eq:Fvp_evol} are
first-order nonlinear ODEs whose right-hand sides depend implicitly on the
Cauchy stress at the current time $t_{n+1}$, which is itself a function of
the unknown internal variables.
Two numerical difficulties arise.
First, a naive explicit update of the form
$\bar{\mathbf{F}}^v_{n+1} = \bar{\mathbf{F}}^v_n + \Delta t\,\dot{\bar{\mathbf{F}}}^v$
does not preserve the isochoric constraints
$\det(\bar{\mathbf{F}}^v) = \det(\bar{\mathbf{F}}^{\text{vp}}) = 1$.
Second, the implicit coupling between the flow directions and the stress
requires an iterative solution within each time step.

Both issues are resolved by an exponential map integrator~\cite{simo1992algorithms}.
For either internal gradient $\bar{\mathbf{F}}^v$ or $\bar{\mathbf{F}}^{\text{vp}}$,
the update at time $t_{n+1}$ reads
\begin{equation}
	\bar{\mathbf{F}}_{n+1}
	= \exp\!\bigl(\Delta t\,\bar{\mathbf{L}}_{n+1}\bigr)\,\bar{\mathbf{F}}_n.
	\label{eq:exp_map}
\end{equation}
Since the flow rules enforce $\operatorname{tr}(\bar{\mathbf{L}}^v) =
\operatorname{tr}(\bar{\mathbf{L}}^{\text{vp}}) = 0$, the identity
$\det(\exp(\Delta t\,\bar{\mathbf{L}})) =
\exp(\Delta t\,\operatorname{tr}\bar{\mathbf{L}}) = 1$
guarantees exact preservation of the isochoric constraint at every iteration.
Because $\bar{\mathbf{L}}_{n+1}$ depends on the unknown stress at $t_{n+1}$,
the update is solved by fixed-point iteration as summarized in
Algorithm~\ref{alg:time_integration}.
The elastic rotation $\mathbf{R}_e$ required by the algorithm is obtained
from the polar decomposition
\begin{equation}
	\bar{\mathbf{F}}^e = \mathbf{R}_e\,\mathbf{U}_e,
	\label{eq:polar_dec}
\end{equation}
where $\mathbf{R}_e$ is proper orthogonal and $\mathbf{U}_e$ is symmetric
positive definite.

\begin{algorithm}[H]
	\caption{Implicit time integration over $[t_n,t_{n+1}]$ with fixed-point
		iteration.}
	\label{alg:time_integration}
	\begin{algorithmic}[1]
		\State \textbf{Given:} $\mathbf{F}_{n+1}$, $\bar{\mathbf{F}}^v_n$,
		$\bar{\mathbf{F}}^{\text{vp}}_n$, $\mathcal{H}_n$, time step $\Delta t$.
		\State \textbf{Initialize:}
		$\bar{\mathbf{F}}^v = \bar{\mathbf{F}}^v_n$,
		$\bar{\mathbf{F}}^{\text{vp}} = \bar{\mathbf{F}}^{\text{vp}}_n$,
		tolerance $\varepsilon_{\text{tol}}$,
		\texttt{converged} $= \text{false}$.
		\While{\texttt{converged} $= \text{false}$}
		\State \textbf{Kinematics:}
		$J = \det(\mathbf{F}_{n+1})$,\;
		$\bar{\mathbf{F}} = J^{-1/3}\mathbf{F}_{n+1}$,\;
		$\bar{\mathbf{F}}^{\text{ve}} = \bar{\mathbf{F}}\,(\bar{\mathbf{F}}^{\text{vp}})^{-1}$,\;
		$\bar{\mathbf{F}}^{e} = \bar{\mathbf{F}}^{\text{ve}}\,(\bar{\mathbf{F}}^v)^{-1}$.
		\State \textbf{Stress:} Compute Cauchy stress $\bm{\sigma}$ via the
		constitutive equations of Section~\ref{subsec:Helmholtz}.
		\State \textbf{Polar decomposition:}
		$\bar{\mathbf{F}}^{e} = \mathbf{R}_e\mathbf{U}_e$.
		\State \textbf{Rotated non-equilibrium stress:}
		$\bm{\sigma}_{\mathrm{neq}}' = \mathbf{R}_e^T\,\bm{\sigma}_{\mathrm{neq}}\,\mathbf{R}_e$.
		\State \textbf{Viscous update:}
		$\tau_{\mathrm{neq}} = \|\operatorname{dev}[\bm{\sigma}_{\mathrm{neq}}]\|_F$;\;
		compute $\dot{\varepsilon}^v$ from Eq.~\eqref{eq:eps_dot_v};\;
		$\bar{\mathbf{D}}^v = \dfrac{\dot{\varepsilon}^v}{\tau_{\mathrm{neq}}}
		\operatorname{dev}[\bm{\sigma}_{\mathrm{neq}}']$;\;
		$\bar{\mathbf{F}}^v_{\text{new}} = \exp\!\bigl(\Delta t\,\bar{\mathbf{D}}^v\bigr)
		\bar{\mathbf{F}}^v$.
		\State \textbf{Viscoplastic update:}
		$\tau_{\mathrm{tot}} = \|\operatorname{dev}[\bm{\sigma}]\|_F$;\;
		\If{$\tau_{\mathrm{tot}} \geq \sigma_0$}
		\State compute $\dot{\varepsilon}^{\text{vp}}$ from Eq.~\eqref{eq:eps_dot_vp};\;
		$\bar{\mathbf{D}}^{\text{vp}} = \dfrac{\dot{\varepsilon}^{\text{vp}}}{\tau_{\mathrm{tot}}}
		\operatorname{dev}[\bm{\sigma}]$;\;
		$\bar{\mathbf{F}}^{\text{vp}}_{\text{new}} =
		\exp\!\bigl(\Delta t\,\bar{\mathbf{D}}^{\text{vp}}\bigr)\bar{\mathbf{F}}^{\text{vp}}$.
		\Else\; $\bar{\mathbf{F}}^{\text{vp}}_{\text{new}} = \bar{\mathbf{F}}^{\text{vp}}$.
		\EndIf
		\State \textbf{Convergence:}
		$\Delta^v = \|\bar{\mathbf{F}}^v_{\text{new}} - \bar{\mathbf{F}}^v\|_F$,\;
		$\Delta^{\text{vp}} = \|\bar{\mathbf{F}}^{\text{vp}}_{\text{new}} -
		\bar{\mathbf{F}}^{\text{vp}}\|_F$.
		\If{$\Delta^v + \Delta^{\text{vp}} < \varepsilon_{\text{tol}}$}
		\State $\bar{\mathbf{F}}^v_{n+1} \leftarrow \bar{\mathbf{F}}^v_{\text{new}}$,\;
		$\bar{\mathbf{F}}^{\text{vp}}_{n+1} \leftarrow
		\bar{\mathbf{F}}^{\text{vp}}_{\text{new}}$,\;
		\texttt{converged} $= \text{true}$.
		\Else\; $\bar{\mathbf{F}}^v \leftarrow \bar{\mathbf{F}}^v_{\text{new}}$,\;
		$\bar{\mathbf{F}}^{\text{vp}} \leftarrow \bar{\mathbf{F}}^{\text{vp}}_{\text{new}}$.
		\EndIf
		\EndWhile
	\end{algorithmic}
\end{algorithm}

The non-equilibrium stress is rotated to the intermediate configuration
to ensure objectivity of the viscous flow rule~\cite{reese1998theory}:
\begin{equation}
	\bm{\sigma}'_{\mathrm{neq}}
	= \mathbf{R}_{e}^{T}\,\bm{\sigma}_{\mathrm{neq}}\,\mathbf{R}_{e}.
	\label{eq:rotated_neq}
\end{equation}
Upon convergence the algorithm satisfies
$\det(\bar{\mathbf{F}}^{v}_{n+1}) = \det(\bar{\mathbf{F}}^{\text{vp}}_{n+1}) = 1$
exactly and delivers a stress state fully consistent with both evolution
equations at $t_{n+1}$.

\subsection{Helmholtz free energy and stress response}
\label{subsec:Helmholtz}

The Helmholtz free energy density is additively decomposed into equilibrium,
non-equilibrium, and volumetric parts~\cite{ambati2016phase}:
\begin{equation}
	\rho_0\psi\left(\bar{\mathbf{B}}^\text{ve}, \bar{\mathbf{B}}^\text{e}, J, \phi\right)
	= g(\phi)\left[\psi_\text{eq}(\bar{\mathbf{B}}^\text{ve})
	+ \psi_\text{neq}(\bar{\mathbf{B}}^\text{e})\right]
	+ g(\phi)\langle\psi_\text{vol}(J)\rangle_+
	+ \langle\psi_\text{vol}(J)\rangle_-,
	\label{eq:free_energy}
\end{equation}
where $g(\phi)$ is the energetic degradation function coupling the free energy to the phase-field variable $\phi \in [0,1]$. $\langle\cdot\rangle_+ = \psi_\text{vol}$ if $J \geq 1$ and zero otherwise, and $\langle\cdot\rangle_- = \psi_\text{vol}$ if $J < 1$ and zero otherwise. This tension--compression asymmetry prevents spurious crack nucleation under compressive loading while retaining both equilibrium and non-equilibrium deviatoric contributions as drivers of fracture. Physical consistency requires
\begin{equation}
	g(0) = 1, \quad g(1) = 0, \quad g'(\phi) \leq 0, \quad g'(1) = 0.
	\label{eq:degradation_conditions}
\end{equation}
The standard quadratic form satisfying these conditions is adopted:
\begin{equation}
	g(\phi) = (1 - \phi)^2 + k,
	\label{eq:degradation_function}
\end{equation}
where $k \ll 1$ is a small residual stiffness parameter introduced to
prevent ill-conditioning at fully damaged integration
points~\cite{miehe2010thermodynamically}.

The equilibrium and non-equilibrium free energies account for contributions
from both the epoxy matrix and from each of the $N_f$ carbon fiber families.
Following~\cite{arash2019viscoelastic2}, these are written as weighted sums
over the fiber families:
\begin{equation}
	\rho_0\psi_{\text{eq}}
	= \sum_{i=1}^{N_f} v_f^{(i)}
	\!\left[\psi_{\text{eq}}^{\text{matrix}}(\bar{\mathbf{B}}^{\text{ve}})
	+ \psi_{\text{eq}}^{\text{fiber}(i)}
	(\bar{\mathbf{C}}^{\text{ve}}, \mathbf{a}_0^{(i)})\right],
	\label{eq:psi_eq}
\end{equation}
\begin{equation}
	\rho_0\psi_{\text{neq}}
	= \sum_{i=1}^{N_f} v_f^{(i)}
	\!\left[\psi_{\text{neq}}^{\text{matrix}}(\bar{\mathbf{B}}^e)
	+ \psi_{\text{neq}}^{\text{fiber}(i)}
	(\bar{\mathbf{C}}^e, \mathbf{a}_0^{(i)})\right],
	\label{eq:psi_neq}
\end{equation}
where $v_f^{(i)}$ is the volume fraction of the $i$-th fiber family,
$\mathbf{a}_0^{(i)}$ is the unit reference fiber direction,
$\bar{\mathbf{C}}^{\text{ve}} = (\bar{\mathbf{F}}^{\text{ve}})^T\bar{\mathbf{F}}^{\text{ve}}$
and $\bar{\mathbf{C}}^e = (\bar{\mathbf{F}}^e)^T\bar{\mathbf{F}}^e$ are the
right Cauchy--Green tensors on each intermediate configuration, and
$\rho_0$ is the reference mass density. The epoxy matrix is modeled by a neo-Hookean hyperelastic potential:
\begin{equation}
	\psi_{\text{eq/neq}}^{\text{matrix}}
	= \frac{1}{2}\mu_{\text{eq/neq}}
	\!\left(\operatorname{tr}[\bar{\mathbf{B}}^{\text{ve/e}}] - 3\right),
	\label{eq:psi_matrix}
\end{equation}
where $\mu_{\text{eq}}$ and $\mu_{\text{neq}}$ are respectively the equilibrium and
non-equilibrium shear moduli of the matrix, captured using a modified Kitagawa model~\cite{unger2020effect}:
\begin{align}
	\mu_{\text{eq}}(\theta) &= \mu_{\text{eq}}^0[2 - \exp(\alpha_{\theta}(\theta - \theta_0))], \label{eq:mu_eq}\\
	\mu_{\text{neq}}(\theta) &= \mu_{\text{neq}}^0[2 - \exp(\alpha_{\theta}(\theta - \theta_0))], \label{eq:mu_neq}
\end{align}
where $\mu_{\text{eq}}^0$ and $\mu_{\text{neq}}^0$ are reference shear moduli at room temperature $\theta_0$ and dry condition, and $\alpha_{\theta}$ is a temperature sensitivity parameter.

The carbon fiber contribution for each family is described by a strain energy
function that captures the coupled response of fiber stretching and
fiber--matrix shear interaction~\cite{arash2019viscoelastic2}:
\begin{align}
	\psi_{\text{fiber}}^{(i)}
	= \frac{1}{2}\mu\bigg[&
	\!\left(v_m + v_f^{(i)}f\!\left(\bar{I}_4^{(i)}\right)\right)
	\!\left(\bar{I}_4^{(i)} + 2\left(\bar{I}_4^{(i)}\right)^{-1/2} - 3\right)
	\nonumber\\
	&+ g_1\!\left(\bar{I}_4^{(i)}\right)
	\!\left(\bar{I}_5^{(i)} - \left(\bar{I}_4^{(i)}\right)^2\right)
	\!\left(\bar{I}_4^{(i)}\right)^{-1}
	\nonumber\\
	&+ g_2\!\left(\bar{I}_4^{(i)}\right)
	\!\left(\bar{I}_1 -
	\!\left(\bar{I}_5^{(i)} + 2\left(\bar{I}_4^{(i)}\right)^{1/2}\right)
	\!\left(\bar{I}_4^{(i)}\right)^{-1}\right)\bigg],
	\label{eq:psi_fiber}
\end{align}
where $v_m = 1 - \sum_{i=1}^{N_f} v_f^{(i)}$ is the matrix volume fraction,
$\mu$ stands for either $\mu_{\text{eq}}$ or $\mu_{\text{neq}}$ as appropriate,
and the fiber invariants are
\begin{align}
	\bar{I}_4^{(i)} &= \mathbf{a}_0^{(i)} \cdot \bar{\mathbf{C}}\,\mathbf{a}_0^{(i)},
	\label{eq:I4}\\
	\bar{I}_5^{(i)} &= \mathbf{a}_0^{(i)} \cdot \bar{\mathbf{C}}^2\mathbf{a}_0^{(i)},
	\label{eq:I5}
\end{align}
representing the squared fiber stretch and a higher-order kinematic
invariant, respectively. The stiffening functions $f$, $g_1$, and $g_2$
in Eq.~\eqref{eq:psi_fiber} are:
\begin{equation}
	f(\bar{I}_4) = a_1 + a_2\exp\!\left[a_3(\bar{I}_4 - 1)\right],
	\label{eq:f_function}
\end{equation}
\begin{equation}
	g_1(\bar{I}_4) =
	\frac{(1 + v_f^{(i)})f(\bar{I}_4) + (1 - v_f^{(i)})}
	{(1 - v_f^{(i)})f(\bar{I}_4) + 1 + v_f^{(i)}},
	\label{eq:g1_function}
\end{equation}
\begin{equation}
	g_2(\bar{I}_4) =
	\frac{(1 + 0.4v_f^{(i)})f(\bar{I}_4) + 0.4(1 - v_f^{(i)})}
	{(1 - v_f^{(i)})f(\bar{I}_4) + 0.4 + v_f^{(i)}},
	\label{eq:g2_function}
\end{equation}
where $a_1$, $a_2$, and $a_3$ are material constants calibrated from
uniaxial tension data of the carbon fiber/epoxy system.

The volumetric free energy is defined as
\begin{equation}
	\rho_0\psi_{\text{vol}}
	= \frac{1}{2}k_v\!\left(\frac{J_m^2 - 1}{2} - \ln J_m\right),
	\label{eq:psi_vol}
\end{equation}
with the bulk modulus $k_v = k_v^0$ taken as the reference value at room temperature.

It should be noted that the left Cauchy--Green tensor $\bar{\mathbf{B}}$ is used for the isotropic
matrix contributions (defined in the current configuration), while the right Cauchy--Green tensor $\bar{\mathbf{C}}$ is employed for the fiber contributions because the fiber invariants in Eqs.~\eqref{eq:I4} and \eqref{eq:I5} are expressed in the reference configuration.

The total Cauchy stress follows from the free energy by standard thermodynamic arguments:
\begin{equation}
	\boldsymbol{\sigma} = g(\phi)\left(
	\boldsymbol{\sigma}_\text{dev}
	+ \langle\boldsymbol{\sigma}_\text{vol}\rangle_+
	\right)
	+ \langle\boldsymbol{\sigma}_\text{vol}\rangle_-,
	\label{eq:sigma_total}
\end{equation}
where $\langle\boldsymbol{\sigma}_\text{vol}\rangle_+$ is the volumetric stress when $J \geq 1$ and zero otherwise, and
$\langle\boldsymbol{\sigma}_\text{vol}\rangle_-$ is the volumetric stress when $J < 1$ and zero otherwise. Also, the deviatoric and volumetric contributions are given by
\begin{equation}
	\boldsymbol{\sigma}_{\text{dev}}
	= J^{-1}\sum_{i=1}^{N_f} v_f^{(i)}
	\!\left[\boldsymbol{\sigma}_{\text{matrix}} + \boldsymbol{\sigma}_{\text{fiber}}^{(i)}\right],
	\label{eq:sigma_dev}
\end{equation}
\begin{equation}
	\boldsymbol{\sigma}_{\text{vol}}
	= \frac{1}{2}k_v J^{-1}\!\left(J_m - \frac{1}{J_m}\right)\mathbf{I}.
	\label{eq:sigma_vol}
\end{equation}

The matrix stress is
\begin{equation}
	\boldsymbol{\sigma}_{\text{matrix}}
	= \mu_{\text{eq}}\operatorname{dev}[\bar{\mathbf{B}}^{\text{ve}}]
	+ \mu_{\text{neq}}\operatorname{dev}[\bar{\mathbf{B}}^e].
	\label{eq:sigma_matrix}
\end{equation}

The fiber stress for each family is derived from the strain energy
function~\eqref{eq:psi_fiber}:
\begin{align}
	\boldsymbol{\sigma}_{\text{fiber}}^{(i)}
	= \frac{2}{J}\bigg[&
	W_1^{(i)}\operatorname{dev}[\bar{\mathbf{B}}]
	+ W_4^{(i)}\bar{I}_4^{(i)}
	\!\left(\mathbf{a} \otimes \mathbf{a} - \tfrac{1}{3}\mathbf{I}\right)
	\nonumber\\
	&+ W_5^{(i)}\bar{I}_4^{(i)}
	\!\left(\mathbf{a}\otimes\bar{\mathbf{B}}\mathbf{a}
	+ \bar{\mathbf{B}}\mathbf{a}\otimes\mathbf{a}
	- \tfrac{2}{3}\bar{I}_5^{(i)}\mathbf{I}\right)\bigg],
	\label{eq:sigma_fiber}
\end{align}
where $\mathbf{a} = \bar{\mathbf{F}}\mathbf{a}_0^{(i)}/(J^{2/3}\bar{I}_4^{(i)})^{1/2}$
is the unit fiber direction in the current configuration, and the scalar
coefficients are
\begin{equation}
	W_1^{(i)} = \frac{1}{2}\mu\, g_2,
	\label{eq:W1}
\end{equation}
\begin{align}
	W_4^{(i)} = \frac{1}{2}\mu\bigg[&
	v_f^{(i)}f'\!\left(\bar{I}_4 + 2\bar{I}_4^{-1/2} - 3\right)
	+ (v_m + v_f^{(i)}f)(1 - \bar{I}_4^{-3/2})
	\nonumber\\
	&- g_1(\bar{I}_5\bar{I}_4^{-2} + 1)
	+ g_2(\bar{I}_5\bar{I}_4^{-2} + \bar{I}_4^{-3/2})
	\nonumber\\
	&+ \frac{\bar{I}_5 - \bar{I}_4^2}{2\bar{I}_4}g_1'
	+ \frac{1}{2}\!\left(\bar{I}_1
	- \frac{\bar{I}_5 + 2\bar{I}_4^{1/2}}{\bar{I}_4}\right)g_2'\bigg],
	\label{eq:W4}
\end{align}
\begin{equation}
	W_5^{(i)} = \frac{\mu}{2\bar{I}_4}(g_1 - g_2),
	\label{eq:W5}
\end{equation}
where primes denote derivatives with respect to $\bar{I}_4^{(i)}$.

\subsection{Fiber orientation and principal fiber families}
\label{subsec:orientation}

In SCFRP, fibers produced by injection or compression
molding are distributed with a range of orientations that depend on the
flow conditions during processing.
A computationally efficient representation is achieved by characterizing the
orientation microstructure through a second-order orientation
tensor:
\begin{equation}
	\mathbf{A} = \frac{1}{N_{\text{fibers}}} \sum_{k=1}^{N_{\text{fibers}}} \mathbf{a}_k \otimes \mathbf{a}_k,
	\label{eq:orientation_tensor}
\end{equation}
where $\mathbf{a}_k$ is the unit direction of the $k$-th fiber and $N_{\text{fibers}}$ is the total number of fibers. The tensor $\mathbf{A}$ is symmetric and positive semi-definite, and its eigenvalues encode the degree of alignment. 
Eigenvalue decomposition of $\mathbf{A}$ yields
\begin{equation}
	\mathbf{A} = \sum_{i=1}^{n_{\text{dim}}} \lambda_i \, \mathbf{n}_i \otimes \mathbf{n}_i,
	\label{eq:eigendecomposition}
\end{equation}
where $\lambda_i$ and $\mathbf{n}_i$ are the eigenvalues and orthonormal
eigenvectors, and $n_{\text{dim}}$ is the spatial dimension.
The eigenvalues satisfy $\sum_i \lambda_i = 1$ and represent the relative
concentration of fibers along each principal direction.
This decomposition maps the continuous orientation distribution onto a small
set of discrete fiber families: the principal direction of the $i$-th family
is $\mathbf{a}_0^{(i)} = \mathbf{n}_i$ and its effective volume fraction is
\begin{equation}
	v_f^{(i)} = v_f \cdot \frac{\lambda_i}{\sum_{j=1}^{N_f} \lambda_j},
	\label{eq:fiber_volume_fraction}
\end{equation}
which ensures conservation of the total fiber content:
\begin{equation}
	\sum_{i=1}^{N_f} v_f^{(i)} = v_f,
	\label{eq:volume_conservation}
\end{equation}
and the matrix volume fraction follows as
\begin{equation}
	v_m = 1 - v_f.
	\label{eq:matrix_volume_fraction}
\end{equation}

The orientation tensor $\mathbf{A}$ serves two purposes in the present
framework.
In the mechanical constitutive model it determines the principal fiber
directions and their weight fractions entering the free energy
expressions~\eqref{eq:psi_eq}--\eqref{eq:psi_neq}.
In the anisotropic phase-field formulation it enters the gradient energy
tensor $\hat{\mathbf{A}}$ controlling the directional crack resistance
(Section~\ref{sec:phase}).
Crucially, the same tensor also defines the anisotropic conductivity of the
carbon fiber network, making the orientation decomposition the common thread
connecting the mechanical, fracture, and electrical sub-problems
(Section~\ref{sec:elec}).

\section{Phase-field model at finite deformation}
\label{sec:phase}

The constitutive model of Section~\ref{sec:const} is embedded within a
variational phase-field fracture framework to predict crack nucleation and
propagation in SCFRP composites.
This section presents the governing equations for the coupled
displacement--phase-field problem at finite deformation, the energy-based
crack driving force, the weak form and finite element (FE) discretization,
and the numerical scheme for the consistent spatial tangent modulus.
The electric field governing equation, which completes the three-field
multiphysics problem, is introduced in Section~\ref{sec:elec}.

\subsection{Governing equations}

Let $\Omega_t \subset \mathbb{R}^{n_{\text{dim}}}$ be the deformed body at
time $t$ with boundary $\Gamma_t$.
The strong form of the coupled boundary value problem for the displacement
field $\mathbf{u}$ and phase-field variable $\phi \in [0,1]$ reads
\begin{align}
	\nabla_{\mathbf{x}} \cdot \boldsymbol{\sigma} + \mathbf{b}
	&= \mathbf{0} \quad \text{in } \Omega_t,
	\label{eq:momentum_spat}\\
	\boldsymbol{\sigma} \cdot \mathbf{n}
	&= \bar{\mathbf{t}} \quad \text{on } \Gamma_t,
	\label{eq:traction_spat}\\
	\frac{G_c}{l_0}\phi
	- G_c l_0 \nabla_{\mathbf{x}} \cdot (\hat{\mathbf{A}} \cdot \nabla_{\mathbf{x}}\phi)
	&= -g'(\phi)\mathcal{H} \quad \text{in } \Omega_t,
	\label{eq:phasefield_spat}\\
	\nabla_{\mathbf{x}}\phi \cdot \mathbf{n}
	&= 0 \quad \text{on } \Gamma_t,
	\label{eq:phasebc_spat}
\end{align}
where $\boldsymbol{\sigma}$ is the Cauchy stress given by
Eq.~\eqref{eq:sigma_total}, $\mathbf{b}$ is the body force per unit
current volume, $\mathbf{n}$ is the outward unit normal on $\Gamma_t$,
$\bar{\mathbf{t}}$ is the prescribed traction, $G_c$ is the critical
energy release rate, and $l_0$ is the length scale parameter governing
the width of the diffuse crack band.

The tensor $\hat{\mathbf{A}}$ in Eq.~\eqref{eq:phasefield_spat} introduces
directional dependence into the crack resistance and is defined
as~\cite{wu2020phase}
\begin{equation}
	\hat{\mathbf{A}} = \mathbf{I} + \hat{\alpha} \mathbf{A},
	\label{eq:aniso_tensor}
\end{equation}
where $\mathbf{A}$ is the fiber orientation tensor of
Eq.~\eqref{eq:orientation_tensor} and $\hat{\alpha} \geq 0$ is a
dimensionless anisotropy parameter.
Setting $\hat{\alpha} = 0$ recovers the isotropic phase-field model.
For $\hat{\alpha} > 0$, the gradient energy term in
Eq.~\eqref{eq:phasefield_spat} penalizes crack gradients along fiber
directions more strongly than across them, so that crack propagation
perpendicular to the fiber axis — which must sever the load-bearing fibers —
requires more energy than propagation parallel to them.
The magnitude of the penalty grows with $\hat{\alpha}$ and with the
eigenvalues of $\mathbf{A}$, i.e., with the degree of fiber alignment.
The critical energy release rate $G_c$ represents the energy required to
create a unit area of crack surface and is calibrated from experimental
fracture tests on the CF/epoxy system.

\subsection{Energy-based crack driving force}
\label{sec:crack_driving_force}

A history-field approach is adopted to account for the irreversibility of
crack growth and to incorporate both elastic and viscous energy contributions
into the fracture driving force~\cite{miehe2010thermodynamically}.
The crack driving force $\mathcal{H}$ is defined as the maximum strain
energy density reached over the entire loading history:
\begin{equation}
	\mathcal{H}(t) = \max_{\tau \in [0,t]} \mathcal{Y}(\tau),
	\label{eq:history_field}
\end{equation}
where the instantaneous energy density available to drive fracture is
\begin{equation}
	\mathcal{Y}
	= \psi_{\text{eq}}(\bar{\mathbf{B}}^{\text{ve}})
	+ \psi_{\text{neq}}(\bar{\mathbf{B}}^e)
	+ \langle\psi_{\text{vol}}(J)\rangle_+,
	\label{eq:energy_density}
\end{equation}
with the positive-part operator $\langle \cdot \rangle_+ = (\cdot + |\cdot|)/2$.
The volumetric contribution is admitted only in tension:
\begin{equation}
	\langle\psi_{\text{vol}}(J)\rangle_+ =
	\begin{cases}
		\psi_{\text{vol}}(J) & \text{if } J \geq 1, \\
		0 & \text{if } J < 1,
	\end{cases}
	\label{eq:vol_split}
\end{equation}
where $J \geq 1$ signifies volume expansion (tension) and $J < 1$ volume
contraction (compression).
This split prevents spurious crack growth under compressive loading while
retaining both equilibrium and non-equilibrium free energy contributions
as drivers of fracture — a physically important feature for
viscoelastic-viscoplastic materials in which viscous energy storage can
represent a significant fraction of the total stored energy.

The history variable is updated at each time step as
\begin{equation}
	\mathcal{H}^{n+1} = \max\!\left(\mathcal{H}^n,\, \mathcal{Y}^{n+1}\right),
	\label{eq:history_update}
\end{equation}
ensuring that $\phi$ can only increase monotonically, consistent with
thermodynamic irreversibility of fracture.

\subsection{Weak form and finite element discretization}

Multiplying Eqs.~\eqref{eq:momentum_spat} and~\eqref{eq:phasefield_spat}
by admissible test functions $\boldsymbol{\eta}_u \in H^1_0(\Omega)$ and
$\eta_\phi \in H^1_0(\Omega)$, integrating over $\Omega_t$, and applying
the divergence theorem yields the weak form:
\begin{equation}
	\int_{\Omega_t} \boldsymbol{\sigma} : \nabla_{\mathbf{x}}\boldsymbol{\eta}_u
	\, dv
	- \int_{\Omega_t} \rho_t \mathbf{b} \cdot \boldsymbol{\eta}_u \, dv
	- \int_{\Gamma_t} \bar{\mathbf{t}} \cdot \boldsymbol{\eta}_u \, da
	= 0,
	\label{eq:weak_u_spat}
\end{equation}
\begin{equation}
	\int_{\Omega_t} \!\left[
	J^{-1}g'(\phi)\mathcal{H}\,\eta_\phi
	+ J^{-1}\frac{G_c}{l_0}\phi\,\eta_\phi
	+ J^{-1}G_c l_0
	\nabla_{\mathbf{x}}\phi \cdot \hat{\mathbf{A}} \cdot \nabla_{\mathbf{x}}\eta_\phi
	\right]\! dv = 0.
	\label{eq:weak_phi_spat}
\end{equation}

\subsection{Consistent incremental-iterative scheme}

Under the assumption of deformation-independent external loading,
Eqs.~\eqref{eq:weak_u_spat} and~\eqref{eq:weak_phi_spat} are cast as
residual equations in terms of internal and external nodal force vectors:
\begin{align}
	\mathbf{r}^u &= \mathbf{f}^u_{\text{int}} - \mathbf{f}^u_{\text{ext}}
	= \mathbf{0},
	\label{eq:residual_u}\\
	\mathbf{r}^\phi &= \mathbf{f}^\phi_{\text{int}} - \mathbf{f}^\phi_{\text{ext}}
	= \mathbf{0},
	\label{eq:residual_phi}
\end{align}
where
\begin{align}
	\mathbf{f}^u_{\text{int}}
	&= \int_{\Omega_t} \boldsymbol{\sigma} : \nabla_x \boldsymbol{\eta}_u \, dv,
	\label{eq:fint_u}\\
	\mathbf{f}^u_{\text{ext}}
	&= \int_{\Omega} \rho_t \mathbf{b} \cdot \boldsymbol{\eta}_u \, dv
	+ \int_{\Gamma_t} \bar{\mathbf{t}} \cdot \boldsymbol{\eta}_u \, da,
	\label{eq:fext_u}\\
	\mathbf{f}^\phi_{\text{int}}
	&= \int_{\Omega_t} \!\left[
	J^{-1}g'(\phi)\mathcal{H}\eta_\phi
	+ J^{-1}\frac{G_c}{l_0}\phi\eta_\phi
	+ J^{-1}G_c l_0
	\nabla_x \phi \cdot \hat{\mathbf{A}} \cdot \nabla_x \eta_\phi
	\right]\! dv,
	\label{eq:fint_phi}\\
	\mathbf{f}^\phi_{\text{ext}}
	&= \mathbf{0}.
	\label{eq:fext_phi}
\end{align}

Newton--Raphson linearization of the residuals at iteration $i+1$ gives
\begin{align}
	\mathbf{r}^u_{i+1} &= \mathbf{r}^u_i + \Delta \mathbf{r}^u = \mathbf{0},
	\label{eq:linearized_u}\\
	\mathbf{r}^\phi_{i+1} &= \mathbf{r}^\phi_i + \Delta \mathbf{r}^\phi
	= \mathbf{0},
	\label{eq:linearized_phi}
\end{align}
where
\begin{align}
	\Delta \mathbf{r}^u
	&= D_u \mathbf{r}^u_i \cdot \Delta \mathbf{u}
	+ D_\phi \mathbf{r}^u_i \cdot \Delta \phi,
	\label{eq:delta_ru}\\
	\Delta \mathbf{r}^\phi
	&= D_u \mathbf{r}^\phi_i \cdot \Delta \mathbf{u}
	+ D_\phi \mathbf{r}^\phi_i \cdot \Delta \phi.
	\label{eq:delta_rphi}
\end{align}

Carrying out the linearization of Eqs.~\eqref{eq:weak_u_spat}
and~\eqref{eq:weak_phi_spat} in the spatial configuration yields
\begin{multline}
	\int_{\Omega_t} \!\left(
	\nabla_x \Delta \mathbf{u} \cdot \boldsymbol{\sigma} \cdot \nabla_x \boldsymbol{\eta}_u
	+ \nabla^s_x \boldsymbol{\eta}_u : \hat{\mathbf{c}} : \nabla^s_x \Delta \mathbf{u}
	\right)\! dv \\
	+ \int_{\Omega_t}
	\nabla^s_x \boldsymbol{\eta}_u : D_\phi \boldsymbol{\sigma} \cdot \Delta \phi
	\, dv
	= \mathbf{f}^u_{\text{ext}} - \mathbf{f}^u_{\text{int},i},
	\label{eq:linearized_momentum}
\end{multline}
and
\begin{equation}
	\begin{split}
		&\int_{\Omega_t} J^{-1}g'(\phi)
		\,2\frac{\partial \mathcal{H}}{\partial \mathbf{g}}
		\cdot \nabla_{\mathbf{x}}\Delta\mathbf{u}
		\;\eta_\phi \, dv \\
		&+ \int_{\Omega_t} \!\left[
		J^{-1}g''(\phi)\mathcal{H}\Delta\phi\,\eta_\phi
		+ J^{-1}\frac{G_c}{l_0}\Delta\phi\,\eta_\phi
		+ J^{-1}G_c l_0
		\nabla_{\mathbf{x}}\Delta\phi \cdot \hat{\mathbf{A}} \cdot \nabla_{\mathbf{x}}\eta_\phi
		\right]\! dv \\
		&= \mathbf{f}^\phi_{\text{ext}} - \mathbf{f}^\phi_{\text{int},i},
	\end{split}
	\label{eq:linearized_phasefield}
\end{equation}
where $\hat{\mathbf{c}}$ is the spatial tangent modulus (the Piola
push-forward of $\partial\mathbf{S}/\partial\mathbf{C}$, with $\mathbf{S}$
the second Piola--Kirchhoff stress), and
$\partial(\cdot)/\partial\mathbf{g} =
\mathbf{F}\,\partial(\cdot)/\partial\mathbf{C}\,\mathbf{F}^T$.

\subsection{Finite element formulation}

The linearized Eqs.~\eqref{eq:linearized_momentum}
and~\eqref{eq:linearized_phasefield} are assembled into the two-field
system
\begin{equation}
	\begin{bmatrix}
		\mathbf{K}^{uu}_i & \mathbf{K}^{u\phi}_i \\
		\mathbf{K}^{\phi u}_i & \mathbf{K}^{\phi\phi}_i
	\end{bmatrix}
	\begin{bmatrix}
		\Delta \mathbf{u}_{i+1} \\
		\Delta \boldsymbol{\phi}_{i+1}
	\end{bmatrix}
	=
	\begin{bmatrix}
		\mathbf{f}^u_{\text{ext}} \\
		\mathbf{f}^\phi_{\text{ext}}
	\end{bmatrix}
	-
	\begin{bmatrix}
		\mathbf{f}^u_{\text{int},i} \\
		\mathbf{f}^\phi_{\text{int},i}
	\end{bmatrix},
	\label{eq:fem_system}
\end{equation}
with the tangent stiffness sub-matrices
\begin{align}
	\mathbf{K}^{uu}_i
	&= \int_{\Omega} \mathbf{B}_u^T \hat{\mathbf{c}}\, \mathbf{B}_u \, d\Omega
	+ \int_{\Omega} \mathbf{B}_u^T \boldsymbol{\sigma}\, \mathbf{B}_u \, d\Omega,
	\label{eq:Kuu}\\
	\mathbf{K}^{u\phi}_i
	&= \int_{\Omega} \mathbf{B}_u^T
	\!\left( \frac{\partial \boldsymbol{\sigma}}{\partial \phi} \right)
	\mathbf{N}_\phi \, d\Omega,
	\label{eq:Kuphi}\\
	\mathbf{K}^{\phi u}_i
	&= \int_{\Omega} \mathbf{N}_\phi^T J^{-1}g'(\phi)
	\,2\frac{\partial \mathcal{H}}{\partial \mathbf{g}} \mathbf{B}_u \, d\Omega,
	\label{eq:Kphiu}\\
	\mathbf{K}^{\phi\phi}_i
	&= \int_{\Omega} J^{-1}\!\left[
	\mathbf{N}_\phi^T \!\left( g''(\phi) \mathcal{H} + \frac{G_c}{l_0} \right)
	\mathbf{N}_\phi
	+ G_c l_0\, \mathbf{B}_\phi^T \hat{\mathbf{A}}\, \mathbf{B}_\phi
	\right]\! d\Omega,
	\label{eq:Kphiphi}
\end{align}
where $\mathbf{N}_u$ and $\mathbf{N}_\phi$ are the shape function matrices
for the displacement and phase-field, and $\mathbf{B}_u$ and $\mathbf{B}_\phi$
are the corresponding gradient operators.
Standard isoparametric Q4 elements with $2 \times 2$ Gauss quadrature are
used for all fields.
The coupled system~\eqref{eq:fem_system} is solved using a staggered
algorithm~\cite{miehe2010phase} in which the displacement and phase-field
sub-problems are solved sequentially within each load step.
After convergence of both sub-problems, the electric field equation
(Section~\ref{sec:elec}) is solved as a decoupled linear system, exploiting
the one-way coupling from the mechanical-damage state to the electrical
response.

\subsection{Consistent tangent modulus based on the Jaumann--Zaremba rate}

At finite deformation the material time derivative of the Cauchy stress is
not objective under superposed rigid body motions.
The Jaumann--Zaremba rate is therefore used to formulate an objective
constitutive tangent:
\begin{equation}
	\overset{\nabla}{\boldsymbol{\sigma}}
	= \dot{\boldsymbol{\sigma}}
	- \mathbf{W}\boldsymbol{\sigma}
	- \boldsymbol{\sigma}\mathbf{W}^T
	= \mathbf{C}^{\sigma J} : \mathbf{D},
	\label{eq:jaumann_rate}
\end{equation}
where $\mathbf{D}$ is the rate of deformation tensor and $\mathbf{W}$ is
the spin tensor.
The spatial tangent $\hat{\mathbf{c}}$ required by the FE weak form in
Eq.~\eqref{eq:Kuu} is the Piola push-forward of the material tangent
$\partial\mathbf{S}/\partial\mathbf{C}$:
\begin{equation}
	\hat{\mathbf{c}} = \frac{1}{J}
	\left(\mathbf{F}\,\overline{\otimes}\,\mathbf{F}\right)
	: \frac{\partial\mathbf{S}}{\partial\mathbf{C}} :
	\left(\mathbf{F}^T\,\overline{\otimes}\,\mathbf{F}^T\right).
	\label{eq:c_hat_pushforward}
\end{equation}

A closed-form evaluation of Eq.~\eqref{eq:c_hat_pushforward} is impractical
for the complex coupled viscoelastic-viscoplastic-damage model.
The Jaumann tangent $\mathbf{C}^{\sigma J}$ is therefore computed numerically
by a forward-difference perturbation scheme~\cite{sun2008numerical}.
A symmetric perturbation is applied to the $(k,l)$ components of the
deformation gradient:
\begin{equation}
	\Delta\mathbf{F}_{kl} = \frac{\epsilon}{2}
	\bigl(\mathbf{e}_k \otimes \mathbf{e}_l\,\mathbf{F}
	+ \mathbf{e}_l \otimes \mathbf{e}_k\,\mathbf{F}\bigr),
	\label{eq:perturbation}
\end{equation}
with $\epsilon = 10^{-5}$.
The corresponding perturbations in the spin and rate-of-deformation tensors
are
\begin{align}
	\Delta\mathbf{W}_{kl} &= \mathbf{0},
	\label{eq:delta_W}\\
	\Delta\mathbf{D}_{kl} &= \frac{\epsilon}{2}
	\bigl(\mathbf{e}_k \otimes \mathbf{e}_l
	+ \mathbf{e}_l \otimes \mathbf{e}_k\bigr).
	\label{eq:delta_D}
\end{align}
Since $\Delta\mathbf{W}_{kl} = \mathbf{0}$, the Jaumann rate reduces to
$\Delta\boldsymbol{\sigma} = \mathbf{C}^{\sigma J}:\Delta\mathbf{D}$,
so the components of $\mathbf{C}^{\sigma J}$ follow from
\begin{align}
	\hat{\mathbf{F}}_{kl}
	&= \mathbf{F} + \Delta\mathbf{F}_{kl},
	\label{eq:F_perturbed}\\
	\Delta\boldsymbol{\sigma}
	&\approx \boldsymbol{\sigma}\!\left(\hat{\mathbf{F}}_{kl}\right)
	- \boldsymbol{\sigma}(\mathbf{F}),
	\label{eq:delta_sigma}\\
	\mathbf{C}^{\sigma J}_{kl}
	&= \frac{1}{\epsilon}
	\Bigl[\boldsymbol{\sigma}\!\left(\hat{\mathbf{F}}_{kl}\right)
	- \boldsymbol{\sigma}(\mathbf{F})\Bigr].
	\label{eq:CsigJ_numerical}
\end{align}

The procedure is summarized in Algorithm~\ref{alg:tangent_algorithm}.

\begin{algorithm}[H]
	\centering
	\caption{Computation of the consistent spatial tangent modulus.}
	\label{alg:tangent_algorithm}
	\begin{tabular}{l}
		\textbf{1.}  Set perturbation parameter $\epsilon = 10^{-5}$. \\
		\textbf{2.}  Evaluate baseline Cauchy stress $\boldsymbol{\sigma}(\mathbf{F})$
		using the constitutive model of Section~\ref{sec:const}. \\
		\textbf{4.}  \textbf{for} $k = 1, \ldots, n_{\text{dim}}$ \textbf{do} \\
		\textbf{5.}  \quad \textbf{for} $l = 1, \ldots, n_{\text{dim}}$ \textbf{do} \\
		\textbf{6.}  \quad\quad Set $\hat{\mathbf{F}} = \mathbf{F}$. \\
		\textbf{7.}  \quad\quad Apply symmetric perturbation:
		$\hat{F}_{kl} \mathrel{+}= \epsilon/2$,\;
		$\hat{F}_{lk} \mathrel{+}= \epsilon/2$. \\
		\textbf{8.}  \quad\quad Evaluate perturbed stress
		$\boldsymbol{\sigma}(\hat{\mathbf{F}})$. \\
		\textbf{9.}  \quad\quad Compute $\Delta\boldsymbol{\sigma}
		= \boldsymbol{\sigma}(\hat{\mathbf{F}}) - \boldsymbol{\sigma}(\mathbf{F})$
		(Eq.~\eqref{eq:delta_sigma}). \\
		\textbf{10.} \quad\quad Store columns $\mathbf{C}^{\sigma J}_{kl}$
		and $\mathbf{C}^{\sigma J}_{lk}$
		using Eq.~\eqref{eq:CsigJ_numerical}. \\
		\textbf{11.} \quad \textbf{end} \\
		\textbf{12.} \textbf{end} \\
		\textbf{13.} Store $\mathbf{C}^{\sigma J}$ in Voigt notation. \\
	\end{tabular}
\end{algorithm}

The spatial tangent $\hat{\mathbf{c}}$ is then recovered from the Jaumann
tangent $\mathbf{C}^{\sigma J}$ by the standard
Jaumann-to-Truesdell correction~\cite{arash2025phase}:
\begin{equation}
	\hat{\mathbf{c}} = \mathbf{C}^{\sigma J}
	- \tfrac{1}{2}\bigl(
	\mathbf{I}\,\overline{\otimes}\,\boldsymbol{\sigma}
	+ \mathbf{I}\,\underline{\otimes}\,\boldsymbol{\sigma}
	+ \boldsymbol{\sigma}\,\overline{\otimes}\,\mathbf{I}
	+ \boldsymbol{\sigma}\,\underline{\otimes}\,\mathbf{I}
	\bigr)
	+ \boldsymbol{\sigma} \otimes \mathbf{I}.
	\label{eq:chat_from_CsigJ}
\end{equation}
The resulting $\hat{\mathbf{c}}$ is symmetrized to remove floating-point
noise, since the theoretical major symmetry of a potential-based tangent is
satisfied exactly only after the Jaumann correction is applied.
This purely numerical differentiation strategy avoids the derivation of
closed-form tangent expressions for the complex coupled constitutive model
while delivering the quadratic convergence rate of the Newton--Raphson
iterations.

\section{Piezoresistive model}
\label{sec:elec}

This section introduces the governing equation for the electric potential,
derives the damage-coupled piezoresistive conductivity tensor, and defines
the resistance-based self-sensing observable. Two physically distinct piezoresistive mechanisms operate simultaneously
under mechanical loading. The first is a geometric-kinematic mechanism. Deformation rotates and
stretches the fibers, altering the effective conduction path lengths and the contact geometry between neighboring fibers. The second is a damage-driven mechanism: matrix cracking and fiber--matrix debonding sever conductive pathways irreversibly, causing resistance to
increase in proportion to the accumulation of phase-field damage $\phi$. The framework developed below captures both mechanisms within a single continuum conductivity tensor.

\subsection{Governing equation for the electric potential}

Under the quasi-static assumption with no free bulk charge, the electric potential $\phi_e$ satisfies the steady-state charge conservation equation in the deformed configuration $\Omega_t$:
\begin{equation}
	\nabla_{\mathbf{x}} \cdot
	\left(\hat{\boldsymbol{\sigma}}_{\text{eff}} \cdot \nabla_{\mathbf{x}}\phi_e\right)
	= 0
	\quad \text{in } \Omega_t,
	\label{eq:laplace}
\end{equation}
with Dirichlet boundary conditions
\begin{equation}
	\phi_e = V_{\text{app}} \;\text{ on } \Gamma_V,
	\qquad
	\phi_e = 0 \;\text{ on } \Gamma_0,
	\label{eq:dbc_elec}
\end{equation}
and homogeneous Neumann conditions
$(\hat{\boldsymbol{\sigma}}_{\text{eff}} \cdot \nabla_{\mathbf{x}}\phi_e)
\cdot \mathbf{n} = 0$ on the remaining boundary. Here $\Gamma_V$ is the voltage electrode, $\Gamma_0$ is the ground electrode, $V_{\text{app}}$ is the applied voltage, and $\hat{\boldsymbol{\sigma}}_{\text{eff}}$ is the effective electrical conductivity tensor derived in Section~\ref{subsec:conductivity}. The electric current density is
\begin{equation}
	\mathbf{J} = -\hat{\boldsymbol{\sigma}}_{\text{eff}} \cdot \nabla_{\mathbf{x}}\phi_e.
	\label{eq:J_density}
\end{equation}

Eq.~\eqref{eq:laplace} is linear in $\phi_e$ for fixed $\hat{\boldsymbol{\sigma}}_{\text{eff}}$, so the electric field sub-problem reduces to a single linear system solve once the mechanical
and phase-field fields have converged.

\subsection{Damage-coupled piezoresistive conductivity tensor}
\label{subsec:conductivity}

The undamaged, unstrained effective conductivity tensor at a material point is assembled by superimposing the isotropic matrix contribution and the transversely isotropic fiber contribution from each principal fiber family $i = 1,\ldots,N_f$:
\begin{equation}
	\hat{\boldsymbol{\sigma}}_0
	= \sigma_m\,\mathbf{I}
	+ \sum_{i=1}^{N_f} v_f^{(i)}
	\!\left[
	\sigma_\parallel^{(i)}\,\mathbf{a}^{(i)}\otimes\mathbf{a}^{(i)}
	+ \sigma_\perp^{(i)}
	\!\left(\mathbf{I} - \mathbf{a}^{(i)}\otimes\mathbf{a}^{(i)}\right)
	\right].
	\label{eq:sigma0_tensor}
\end{equation}

The principal fiber directions $\mathbf{a}^{(i)}$ in Eq.~\eqref{eq:sigma0_tensor} coincide with 
the eigenvectors $\mathbf{n}_i$ of the orientation tensor $\mathbf{A}$ 
(Eq.~\eqref{eq:eigendecomposition}), and the associated volume fractions $v_f^{(i)}$ are 
determined by Eq.~\eqref{eq:fiber_volume_fraction}. Thus, $\hat{\boldsymbol{\sigma}}_0$ 
inherits its anisotropy directly from $\mathbf{A}$.

The tensor structure in Eq.~\eqref{eq:sigma0_tensor} is transversely
isotropic about $\mathbf{a}^{(i)}$. The projection
$\mathbf{a}^{(i)}\otimes\mathbf{a}^{(i)}$ carries conductivity
$\sigma_\parallel^{(i)}$ along the fiber axis, while the complementary
projector $(\mathbf{I} - \mathbf{a}^{(i)}\otimes\mathbf{a}^{(i)})$
carries conductivity $\sigma_\perp^{(i)}$ in all directions transverse
to the fiber.
This representation is fully consistent with the fiber family decomposition
of Section~\ref{subsec:orientation}.

The strain-dependent axial and transverse conductivities follow a linear piezoresistive law~\cite{abry1999situ}:
\begin{align}
	\sigma_\parallel^{(i)}
	&= \sigma_\parallel^0
	\left(1 - \mathsf{GF}_\parallel\,\varepsilon_\parallel^{(i)}\right),
	\label{eq:sigma_par}\\
	\sigma_\perp^{(i)}
	&= \sigma_\perp^0
	\left(1 - \mathsf{GF}_\perp\,\varepsilon_\perp^{(i)}\right),
	\label{eq:sigma_perp}
\end{align}
where $\mathsf{GF}_\parallel$ and $\mathsf{GF}_\perp$ are the longitudinal and transverse gauge factors, respectively. Mechanical deformation changes the intrinsic conductivity of each fiber family. Therefore, the fiber-parallel and fiber-transverse strain scalars for the $i$-th family are
\begin{align}
	\varepsilon_\parallel^{(i)}
	&= \mathbf{a}_0^{(i)} \cdot \mathbf{E}\,\mathbf{a}_0^{(i)},
	\label{eq:eps_par}\\
	\varepsilon_\perp^{(i)}
	&= \frac{1}{2}\!\left(
	\operatorname{tr}[\mathbf{E}] - \varepsilon_\parallel^{(i)}
	\right).
	\label{eq:eps_perp}
\end{align}

Progressive matrix cracking and fiber--matrix debonding sever the conductive carbon fiber network, degrading the macroscopic conductivity in proportion to the phase-field damage variable $\phi$. This is modeled by multiplying the conductivity tensor by a scalar degradation function:
\begin{equation}
	\hat{\boldsymbol{\sigma}}_{\text{eff}}
	= h_e(\phi)\,\hat{\boldsymbol{\sigma}}_0,
	\label{eq:sigma_damage}
\end{equation}
where
\begin{equation}
	h_e(\phi) = (1 - \phi)^p + k_e.
	\label{eq:h_elec}
\end{equation}
The exponent $p$ controls the rate at which conductivity degrades as
damage accumulates, and $k_e$ is a small residual conductivity
retained for numerical stability at fully damaged points.
Setting $p = 2$ reproduces the same quadratic form as the mechanical
degradation function $g(\phi) = (1-\phi)^2 + k$
(Eq.~\eqref{eq:degradation_function}), enforcing consistent degradation kinetics. 

The primary structural health monitoring observable is the specimen-level electrical resistance $R$, obtained from the total current $I$ flowing through the voltage electrode:
\begin{equation}
	I = \int_{\Gamma_V} \mathbf{J} \cdot \mathbf{n}\, da,
	\label{eq:total_current}
\end{equation}
\begin{equation}
	R = \frac{V_{\text{app}}}{I}.
	\label{eq:resistance}
\end{equation}

The normalized conductivity, defined as the ratio of the current resistance $R$ to the initial undamaged resistance $R_0$, provides a dimensionless self-sensing indicator bounded between 0 and 1:
\begin{equation}
	\frac{\sigma}{\sigma_0} = \frac{R_0}{R},
	\label{eq:sigma_norm}
\end{equation}
where $R_0$ is the resistance computed at the first load step before any deformation or damage. The normalized conductivity starts at unity in the undamaged composite and decreases monotonically toward zero as fracture progresses.

\subsection{Weak form and finite element discretization}
\label{subsec:fem_piezo}

Multiplying Eq.~\eqref{eq:laplace} by a test function
$\eta_e \in H^1_0(\Omega_t)$, integrating over $\Omega_t$, and applying
the divergence theorem yields the weak form of the electric field
problem:
\begin{equation}
	\int_{\Omega_t}
	\nabla_{\mathbf{x}}\eta_e
	\cdot \hat{\boldsymbol{\sigma}}_{\text{eff}}
	\cdot \nabla_{\mathbf{x}}\phi_e\,dv
	= 0.
	\label{eq:weak_elec}
\end{equation}
The discretized nodal electric potentials $\boldsymbol{\phi}_e$ are interpolated as
$\phi_e \approx \mathbf{N}_e\boldsymbol{\phi}_e$, where $\mathbf{N}_e$ collects the element shape functions. The element conductivity matrix is
\begin{equation}
	\mathbf{K}^{ee}_e
	= \int_{\Omega_t^e}
	\mathbf{B}_e^T\,\hat{\boldsymbol{\sigma}}_{\text{eff}}\,\mathbf{B}_e\,dv,
	\label{eq:Kee}
\end{equation}
where $\mathbf{B}_e = \nabla_{\mathbf{x}}\mathbf{N}_e$ is the electric
potential gradient matrix evaluated in the current configuration. The conductivity tensor $\hat{\boldsymbol{\sigma}}_{\text{eff}}$ is evaluated at each Gauss point using Eq.~\eqref{eq:sigma_damage} with the converged $\mathbf{F}$ and $\phi$ fields. The global system
$\mathbf{K}^{ee}\boldsymbol{\phi}_e = \mathbf{f}^e$
is linear and is solved in a single step. The element load vector $\mathbf{f}^e_e$ arises solely from the Dirichlet boundary conditions on the electrode surfaces.

The mechanical, phase-field, and electric field sub-problems are coupled in a one-way cascade. Deformation and damage govern the conductivity, but
the electric field does not affect the mechanical or fracture response. This one-way coupling is exploited by solving the three fields sequentially within each load step, as summarized in
Algorithm~\ref{alg:staggered}. 

\begin{algorithm}[H]
	\caption{Staggered solution algorithm for the three-field coupled problem.}
	\label{alg:staggered}
	\begin{algorithmic}[1]
		\State \textbf{Given:}
		displacement increment $\Delta\bar{u}$, $\Delta t$,
		$V_{\text{app}}$, $\varepsilon_{\text{tol}}$,
		max.\ load reductions $n_{\text{red}}$, load reduction factor $k_{\text{red}}$.
		\State Apply $\Delta\bar{u}$; initialize Newton--Raphson counter
		$n_{\text{NR}} = 1$.
		\While{simulation not complete}
		\State \textbf{Displacement sub-problem (Newton--Raphson):}
		\State \quad Assemble $\mathbf{K}^{uu}$ and $\mathbf{f}^u_{\text{int}}$
		(Eqs.~\eqref{eq:Kuu}, \eqref{eq:fint_u}).
		\State \quad Solve $\mathbf{K}^{uu}\Delta\mathbf{u} =
		\mathbf{f}^u_{\text{ext}} - \mathbf{f}^u_{\text{int}}$.
		\State \quad Update $\bar{\mathbf{F}}^v_{n+1}$, $\bar{\mathbf{F}}^{\text{vp}}_{n+1}$
		via Algorithm~\ref{alg:time_integration}.
		\State \textbf{Phase-field sub-problem (Newton--Raphson):}
		\State \quad Assemble $\mathbf{K}^{\phi\phi}$ and $\mathbf{f}^\phi_{\text{int}}$
		(Eqs.~\eqref{eq:Kphiphi}, \eqref{eq:fint_phi}).
		\State \quad Solve $\mathbf{K}^{\phi\phi}\Delta\boldsymbol{\phi} =
		-\mathbf{f}^\phi_{\text{int}}$.
		\State \quad Update history variable:
		$\mathcal{H}^{n+1} = \max(\mathcal{H}^n, \mathcal{Y}^{n+1})$
		(Eq.~\eqref{eq:history_update}).
		\If{$\|\Delta\mathbf{u}\|/\|\mathbf{u}\| \leq \varepsilon_{\text{tol}}$
			\textbf{and}
			$\|\Delta\boldsymbol{\phi}\|/\|\boldsymbol{\phi}\| \leq \varepsilon_{\text{tol}}$}
		\State \textbf{Electric field sub-problem (linear solve):}
		\State \quad Evaluate $\hat{\boldsymbol{\sigma}}_{\text{eff}}$ at all Gauss
		points via Eq.~\eqref{eq:sigma_damage}.
		\State \quad Assemble $\mathbf{K}^{ee}$ (Eq.~\eqref{eq:Kee}).
		\State \quad Apply BCs: $\phi_e = V_{\text{app}}$ on $\Gamma_V$,\;
		$\phi_e = 0$ on $\Gamma_0$.
		\State \quad Solve $\mathbf{K}^{ee}\boldsymbol{\phi}_e = \mathbf{f}^e$.
		\State \quad Compute $\mathbf{J} =
		-\hat{\boldsymbol{\sigma}}_{\text{eff}}\,\nabla_{\mathbf{x}}\phi_e$
		(Eq.~\eqref{eq:J_density}).
		\State \quad Compute $I$ (Eq.~\eqref{eq:total_current}),
		$R = V_{\text{app}}/I$ (Eq.~\eqref{eq:resistance}),
		$\sigma/\sigma_0 = R_0/R$ (Eq.~\eqref{eq:sigma_norm}).
		\State Store $\{u,\,F,\,\phi,\,R,\,\sigma/\sigma_0\}$;
		advance load step.
		\Else
		\If{$n_{\text{NR}} \leq n_{\text{red}}$}
		\State Reduce $\Delta\bar{u} \leftarrow \Delta\bar{u}/k_{\text{red}}$;
		$\Delta t \leftarrow \Delta t/k_{\text{red}}$;
		$n_{\text{NR}} \mathrel{+}= 1$; \textbf{restart step}.
		\Else\; \textbf{terminate}.
		\EndIf
		\EndIf
		\EndWhile
	\end{algorithmic}
\end{algorithm}


\section{Structural health monitoring}
\label{sec:shm_inverse}

The coupled phase-field and piezoresistive framework developed in the
preceding sections provides the physical basis for a non-invasive
structural health monitoring system: as a crack initiates and propagates
under mechanical loading, the spatial distribution of the phase-field
variable $\phi$ degrades the effective conductivity tensor
$\hat{\boldsymbol{\sigma}}_\mathrm{eff}$ through the damage function
$h_e(\phi)$, producing measurable changes in the inter-electrode
conductances that encode the evolving damage state of the specimen.
To exploit this coupling for in-service damage monitoring, an
EIT configuration is
adopted, following the electrode layouts employed in experimental
piezoresistive studies of carbon fiber-reinforced composites~\cite{todoroki2005apparent}.

\begin{figure}[!ht]
	\centering
	\includegraphics{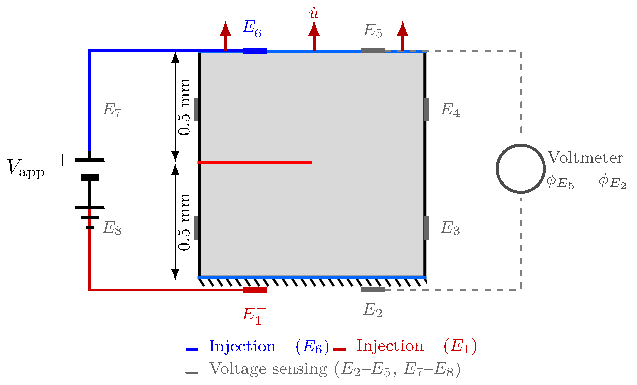}
	\caption{Eight-electrode EIT configuration for spatial conductivity mapping of the single-edge notched specimen. The top edge is subjected to a prescribed tensile displacement
		$\dot{u} = 1\,\mathrm{mm/min}$.}
	\label{fig:eit_electrodes}
\end{figure}

The electrode arrangement is illustrated in Fig.~\ref{fig:eit_electrodes}.
Eight patch electrodes $\mathrm{E}_1$--$\mathrm{E}_8$, each of nominal
half-width $l_c$, are placed at the quarter-points of the four boundary
edges of the single-edge notched specimen, with two electrodes per edge.
For a given injection pair, a constant voltage $V_\mathrm{app}$ is
applied at the anode electrode while the cathode is held at zero
potential; the remaining six electrodes act as passive voltage sensors,
recording the inter-electrode potential differences that reflect the
spatial distribution of $\hat{\boldsymbol{\sigma}}_\mathrm{eff}$ in the
interior.
Cycling sequentially through all $\binom{8}{2} = 28$ injection pairs
and recording the resulting conductance
\begin{equation}
	G_{ij}^{(n)} = \frac{1}{V_\mathrm{app}}
	\left\lvert
	\sum_{k \in \mathcal{E}_i}
	\bigl(\hat{\boldsymbol{\sigma}}_\mathrm{eff}
	\cdot \nabla \phi_e\bigr) \cdot \mathbf{n}\,\mathrm{d}A
	\right\rvert
	\label{eq:conductance}
\end{equation}
at each load step yields 28 independent scalar measurements per step,
assembled into the measurement vector
$\mathbf{g}^{(n)} = \{G_{ij}^{(n)}\}_{i<j} \in \mathbb{R}^{28}$.
The normalised conductance ratios $G_{ij}^{(n)}/G_{ij}^0$, where
$G_{ij}^0$ denotes the undamaged reference conductance measured at the
start of loading, serve as the primary electrical observables: they are
bounded in $(0,\,1]$, monotonically decreasing with damage progression,
and insensitive to the absolute conductivity level, which removes the
dependence on fiber volume fraction and contact resistance variations
between specimens.
The top edge of the specimen is subjected to a prescribed displacement
rate $\dot{u} = 1$~mm/min, consistent with the quasi-static loading
conditions assumed in the constitutive model.

The inverse problem seeks the current damage state from the electrical
measurements alone.
Two scalar global damage indicators are adopted as outputs:
(1) the normalized crack length $\tilde{a} = \ell_\mathrm{crack} / W$,
obtained from the crack surface density functional
\begin{equation}
	\ell_\mathrm{crack}
	= \int_\Omega \left(
	\frac{\phi^2}{2 l_c} + \frac{l_c}{2} |\nabla\phi|^2
	\right) \mathrm{d}V ,
	\label{eq:crack_length}
\end{equation}
and (2) the normalized mechanical compliance
$\tilde{C} = C / C_0 = (u_\mathrm{top} / F_\mathrm{top}) / C_0$,
where $C_0$ is the undamaged reference compliance.
Both quantities are continuous, monotonically increasing functions of the
damage state and are threshold-free by construction, in contrast to
point-based crack tip locators that are sensitive to mesh resolution and
the choice of a phase-field threshold.

The inverse mapping to be learned is therefore
\begin{equation}
	\mathcal{F} \colon
	\left(
	A_{11},\, A_{12},\, v_f,\, \theta,\,
	\mathbf{g}^{(n)} / \mathbf{g}^0
	\right)
	\longmapsto
	\left( \tilde{a}^{(n)},\, \tilde{C}^{(n)} \right),
	\label{eq:inverse_map}
\end{equation}
where $A_{11}$ and $A_{12}$ are the independent components of the
second-order fiber orientation tensor $\mathbf{A}$
(with the constraint $A_{11} + A_{22} = 1$), 
and $\mathbf{g}^{(n)} / \mathbf{g}^0 \in (0, 1]^{28}$ is the vector of
28 normalized conductance ratios at load step $n$.
The microstructure descriptors $(A_{11}, A_{12}, v_f)$ are
fixed properties of the specimen, available from manufacturing data;
the conductance ratios are the sole time-varying measurements.
The total input dimension is therefore $4 + 28 = 32$.

The mapping $\mathcal{F}$ is nonlinear and implicitly defined through the
coupled three-field problem.
Its direct inversion is intractable in closed form,
and iterative forward solves would be computationally prohibitive in a
real-time monitoring context.
The present work approximates $\mathcal{F}$ by a feedforward ANN trained on the phase-field simulation dataset
described in Section~\ref{sec:ann_training}.
The key advantage of the orientation tensor parameterization is that
$\mathbf{A}$ provides a continuous, physically consistent representation
of the microstructure that interpolates smoothly across single-family,
multi-family, and randomly oriented fiber architectures, enabling the
trained model to generalize to configurations not seen during training.

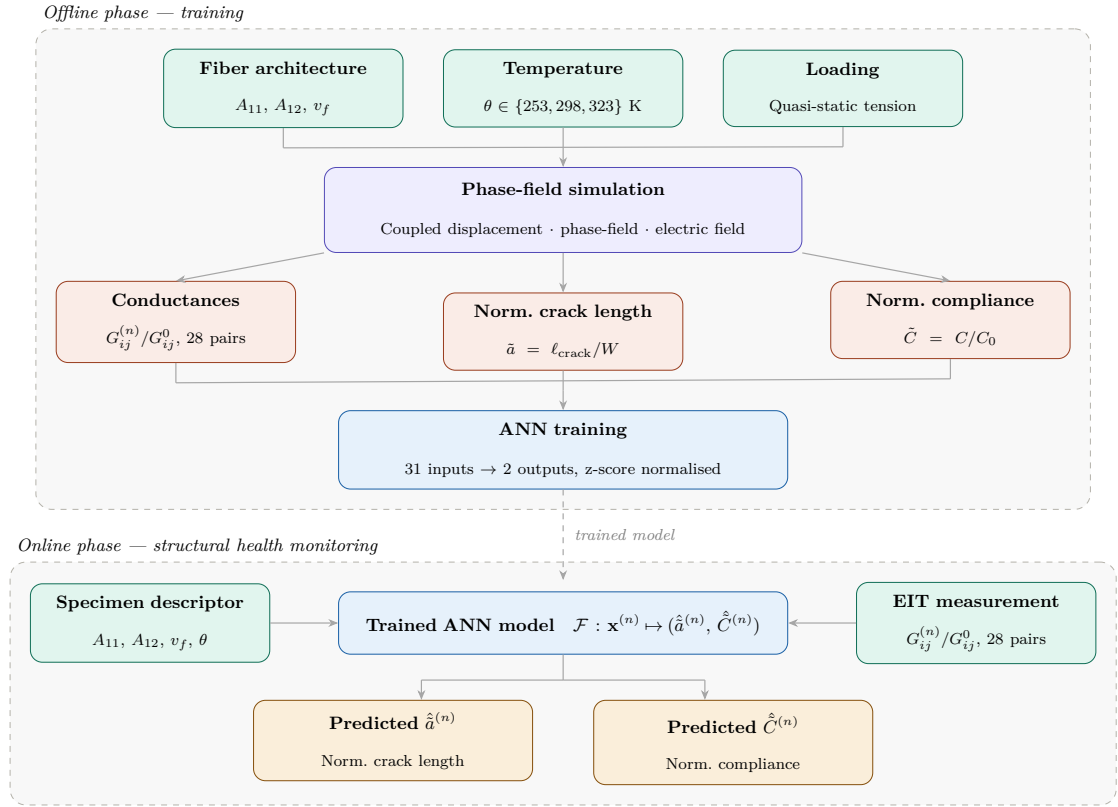
\begin{figure}[H]
\centering
\scalebox{0.75}{%
\begin{tikzpicture}[node distance=0.6cm and 0.5cm]
	
	
	\node[tealbox] (fiber) {%
		\textbf{Fiber architecture}\\[2pt]
		\footnotesize $A_{11}$, $A_{12}$, $v_f$};
	
	\node[tealbox, right=0.7cm of fiber] (temp) {%
		\textbf{Temperature}\\[2pt]
		\footnotesize $\theta\!\in\!\{253,298,323\}$~K};
	
	\node[tealbox, right=0.7cm of temp] (load) {%
		\textbf{Loading}\\[2pt]
		\footnotesize Quasi-static tension};
	
	\coordinate (fmid) at ([yshift=-0.35cm]fiber.south);
	\coordinate (tmid) at ([yshift=-0.35cm]temp.south);
	\coordinate (lmid) at ([yshift=-0.35cm]load.south);
	\coordinate (cmid) at ([yshift=-0.35cm]temp.south);
	
	\draw[thinline] (fiber.south) -- (fmid);
	\draw[thinline] (temp.south)  -- (tmid);
	\draw[thinline] (load.south)  -- (lmid);
	\draw[thinline] (fmid) -- (lmid);
	\draw[arr]      (cmid) -- ++(0,-0.35cm);
	
	\node[purplebox, below=0.7cm of temp] (pf) {%
		\textbf{Phase-field simulation}\\[3pt]
		\footnotesize Coupled displacement $\cdot$ phase-field $\cdot$ electric field};
	
	\node[coralbox, below left=0.7cm of pf] (cond) {%
		\textbf{Conductances}\\[2pt]
		\footnotesize $G_{ij}^{(n)}/G_{ij}^0$, 28 pairs};
	
	\node[coralbox, below=0.7cm of pf] (atilde) {%
		\textbf{Norm.\ crack length}\\[2pt]
		\footnotesize $\tilde{a}=\ell_\mathrm{crack}/W$};
	
	\node[coralbox, below right=0.7cm of pf] (ctilde) {%
		\textbf{Norm.\ compliance}\\[2pt]
		\footnotesize $\tilde{C}=C/C_0$};
	
	\draw[arr] (pf.south)      -- (atilde.north);
	\draw[arr] (pf.south west) -- (cond.north);
	\draw[arr] (pf.south east) -- (ctilde.north);
	
	\coordinate (cmid2)  at ([yshift=-0.35cm]atilde.south);
	\coordinate (cleft)  at ([yshift=-0.35cm]cond.south);
	\coordinate (cright) at ([yshift=-0.35cm]ctilde.south);
	
	\draw[thinline] (cond.south)   -- (cleft);
	\draw[thinline] (atilde.south) -- (cmid2);
	\draw[thinline] (ctilde.south) -- (cright);
	\draw[thinline] (cleft) -- (cright);
	\draw[arr]      (cmid2) -- ++(0,-0.35cm);
	
	\node[bluebox, below=0.7cm of atilde] (ann) {%
		\textbf{ANN training}\\[3pt]
		\footnotesize 31 inputs $\rightarrow$ 2 outputs,
		z-score normalised};
	
	\begin{scope}[on background layer]
		\node[draw=phaseborder, fill=phasebg, dashed, rounded corners=8pt,
		fit=(fiber)(temp)(load)(pf)(cond)(atilde)(ctilde)(ann),
		inner sep=10pt, line width=0.5pt] (offbox) {};
	\end{scope}
	\node[phaselabel, anchor=south west]
	at (offbox.north west) {\textcolor{black}{Offline phase --- training}};
	
	\draw[dasharr] (ann.south) -- ++(0,-1.6cm)
	node[midway, right=2pt, font=\footnotesize\itshape, text=gray]
	{trained model};
	
	
	\node[bluebox, below=1.8cm of ann, minimum height=1.1cm] (model) {%
		\textbf{Trained ANN model}\quad
		$\mathcal{F}:\mathbf{x}^{(n)}\!\mapsto\!
		(\hat{\tilde{a}}^{(n)},\,\hat{\tilde{C}}^{(n)})$};
	
	\node[tealbox, left=1.2cm of model] (spec) {%
		\textbf{Specimen descriptor}\\[2pt]
		\footnotesize $A_{11}$, $A_{12}$, $v_f$, $\theta$};
	
	\node[tealbox, right=1.2cm of model] (eit) {%
		\textbf{EIT measurement}\\[2pt]
		\footnotesize $G_{ij}^{(n)}/G_{ij}^0$, 28 pairs};
	
	\draw[arr] (spec.east) -- (model.west);
	\draw[arr] (eit.west)  -- (model.east);
	
	\coordinate (modout) at ([yshift=-0.45cm]model.south);
	\draw[thinline] (model.south) -- (modout);
	
	\coordinate (outleft)  at ([xshift=-2.5cm]modout);
	\coordinate (outright) at ([xshift= 2.5cm]modout);
	\draw[thinline] (modout) -- (outleft);
	\draw[thinline] (modout) -- (outright);
	\draw[arr] (outleft)  -- ++(0,-0.35cm);
	\draw[arr] (outright) -- ++(0,-0.35cm);
	
	\node[amberbox, below=0.8cm of model, xshift=-3.0cm] (pred_a) {%
		\textbf{Predicted $\hat{\tilde{a}}^{(n)}$}\\[2pt]
		\footnotesize Norm.\ crack length};
	
	\node[amberbox, below=0.8cm of model, xshift=3.0cm] (pred_C) {%
		\textbf{Predicted $\hat{\tilde{C}}^{(n)}$}\\[2pt]
		\footnotesize Norm.\ compliance};
	
	\begin{scope}[on background layer]
		\node[draw=phaseborder, fill=phasebg, dashed, rounded corners=8pt,
		fit=(model)(spec)(eit)(pred_a)(pred_C),
		inner sep=10pt, line width=0.5pt] (onbox) {};
	\end{scope}
	\node[phaselabel, anchor=south west]
	at (onbox.north west)
	{\textcolor{black}{Online phase --- structural health monitoring}};
		
\end{tikzpicture}
}
\caption{Schematic of the phase-field-informed ANN framework for SHM of SCFRP composites. Offline phase (top): phase-field simulations spanning a range of fiber architectures, temperatures, and loading conditions generate	training data. Online phase (bottom): at each monitoring cycle the trained	model $\mathcal{F}$ receives the fixed specimen descriptor $(A_{11}, A_{12}, v_f)$ and the current EIT conductance	measurements, and returns real-time predictions of the normalized crack	length $\hat{\tilde{a}}^{(n)}$ and compliance $\hat{\tilde{C}}^{(n)}$ without mechanical sensing.}
\label{fig:shm_flowchart}
\end{figure}

The overall workflow of the phase-field-informed SHM framework is summarized in Fig.~\ref{fig:shm_flowchart}. The framework operates in two distinct phases. In the offline training phase, phase-field simulations are executed for a systematic set of fiber architectures, temperatures, and loading conditions, producing per-step datasets of normalized conductances
$G_{ij}^{(n)}/G_{ij}^0$, normalised crack length $\tilde{a}$, and normalized compliance $\tilde{C}$. In the online monitoring phase, the trained model receives two inputs
at each monitoring cycle: the fixed specimen descriptor
$(A_{11}, A_{12}, v_f)$ available from manufacturing data, and the 28 conductance ratios measured electrically from the eight-electrode network. The model then instantly returns estimates of the current normalized crack length $\hat{\tilde{a}}^{(n)}$ and normalized compliance
$\hat{\tilde{C}}^{(n)}$, enabling real-time damage tracking without any mechanical loading or measurement.

\tikzexternaldisable
\tikzexternalenable

\section{Numerical simulations}
\label{sec:results}

\subsection{Phase-field simulations}
\label{sec:phase_results}

A comprehensive set of numerical simulations is presented to demonstrate the predictive
capability of the proposed three-field framework and to elucidate the interplay between
anisotropic viscoelastic-viscoplastic fracture and damage-induced piezoresistive response
in SCFRP composites.  All simulations examine
CF/epoxy composites under quasi-static tensile loading.  The material parameters for the
constitutive model, the phase-field fracture formulation, and the piezoresistive electrical
model are summarized in Table~\ref{tab:param}.  The mechanical and fracture parameters are
calibrated from experimental data reported in the literature for the same composite
system~\cite{BAHTIRI2023116293,arash2023effect}, while the electrical parameters are taken
from measurements on carbon fibers~\cite{ji2022anisotropic}.  The numerical investigations
systematically examine the spatial distribution of the crack driving force under viscous
relaxation for balanced and unbalanced fiber architectures and the coupled electromechanical response. 

The mechanical parameters of the constitutive model were calibrated against experimental force--displacement data reported in~\cite{bahtiri2023machine,arash2023effect} for epoxy composites, establishing quantitative agreement for stiffness, peak load, and post-peak softening. For the piezoresistive sub-model, direct experimental validation is beyond the scope of the present work. However, the predicted electromechanical signature is consistent with the experimental observations for CFRP laminates~\cite{abry1999situ,todoroki2004electrical}.

\begin{figure}[H]
\centering
\begin{subfigure}[b]{0.45\linewidth}
	\centering
	\includegraphics{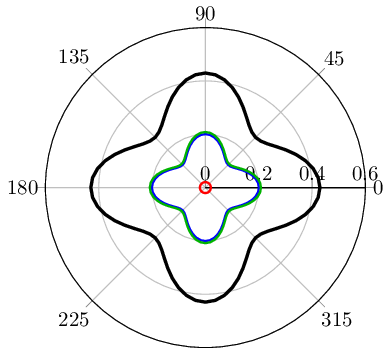}
	\caption{ }
	\label{fig:driving_force0_1}
\end{subfigure}\hspace{24pt}
\begin{subfigure}[b]{0.45\linewidth}
	\centering
	\includegraphics{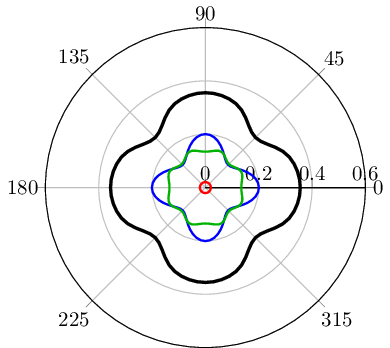}
	\caption{ }
	\label{fig:driving_force0_2}
\end{subfigure}

\begin{subfigure}[b]{0.45\linewidth}
	\centering
	\includegraphics{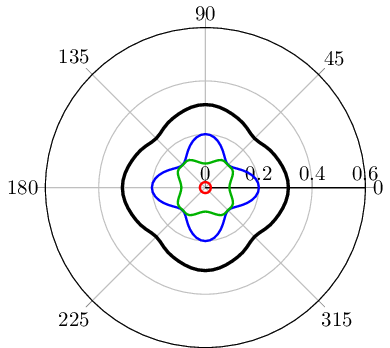}
	\caption{ }
	\label{fig:driving_force0_3}
\end{subfigure}\hspace{24pt}
\begin{subfigure}[b]{0.45\linewidth}
	\centering
	\includegraphics{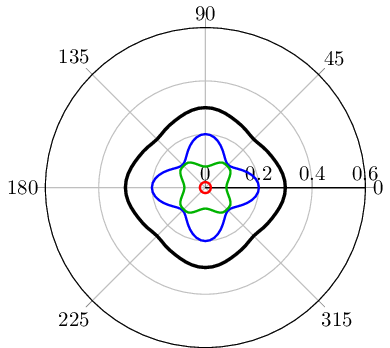}
	\caption{ }
	\label{fig:driving_force0_4}
\end{subfigure}
\caption{Spatial distribution of crack driving force components during viscous relaxation for CF (50 wt\%)/epoxy with balanced fiber families oriented at $\pm 45^{\circ}$ at 296~K. The contours show (a) initial response at $t = 10^{-6}$~s, (b) early relaxation at $t = 0.005$~s, (c) intermediate relaxation at $t = 0.05$~s, and (d) near-equilibrium at $t = 0.1$~s. Color coding: total (black), equilibrium (blue), non-equilibrium (green), and volumetric (red) energy contributions.}
\label{fig:driving_force0}
\end{figure}

\begin{figure}[H]
\centering
\begin{subfigure}[b]{0.45\linewidth}
	\centering
	\includegraphics{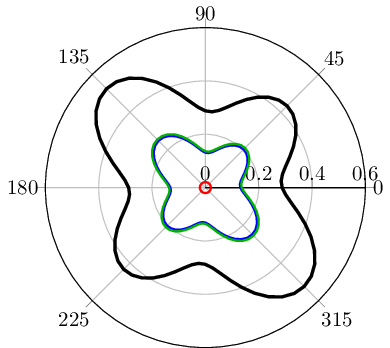}
	\caption{ }
	\label{fig:driving_force1_1}
\end{subfigure}\hspace{24pt}
\begin{subfigure}[b]{0.45\linewidth}
	\centering
	\includegraphics{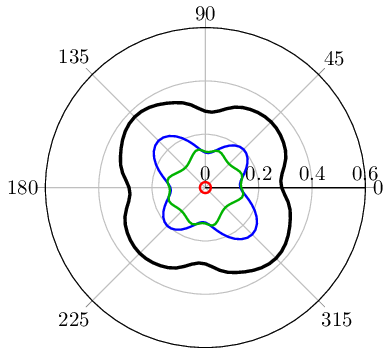}
	\caption{ }
	\label{fig:driving_force1_2}
\end{subfigure}

\begin{subfigure}[b]{0.45\linewidth}
	\centering
	\includegraphics{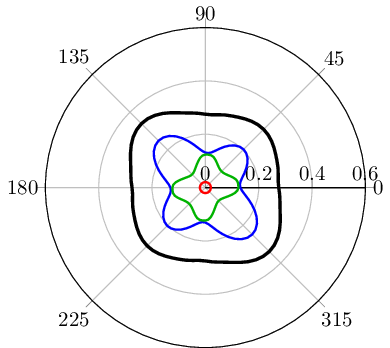}
	\caption{ }
	\label{fig:driving_force1_3}
\end{subfigure}\hspace{24pt}
\begin{subfigure}[b]{0.45\linewidth}
	\centering
	\includegraphics{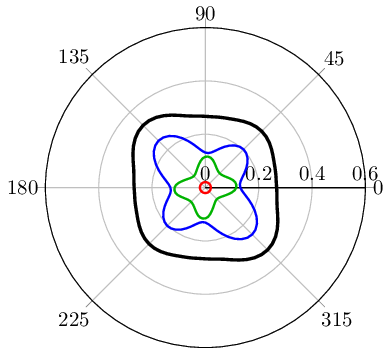}
	\caption{ }
	\label{fig:driving_force1_4}
\end{subfigure}
\caption{Spatial distribution of crack driving force components during viscous relaxation for CF (50 wt\%)/epoxy with unbalanced fiber distribution (70 wt\% at $-45^{\circ}$ and 30 wt\% at $+45^{\circ}$) at 296~K. The contours show (a) initial response at $t = 10^{-6}$~s, (b) early relaxation at $t = 0.005$~s, (c) intermediate relaxation at $t = 0.05$~s, and (d) near-equilibrium at $t = 0.1$~s. Color coding: total (black), equilibrium (blue), non-equilibrium (green), and volumetric (red) energy contributions.}
\label{fig:driving_force}
\end{figure}

We first characterize how viscoelastic and viscoplastic energy storage mechanisms
differentially contribute to the crack driving force in CF/epoxy composites.
Figs.~\ref{fig:driving_force0} and~\ref{fig:driving_force} present polar diagrams of the
instantaneous crack driving force $\mathcal{H}$ at four time instants following an applied
uniaxial strain for CF (50\,wt\%)/epoxy at 296\,K.
Fig.~\ref{fig:driving_force0} examines a balanced distribution with equal fiber weight
fractions at $\pm 45^{\circ}$, while Fig.~\ref{fig:driving_force} presents an unbalanced
configuration with 70\,wt\% at $-45^{\circ}$ and 30\,wt\% at $+45^{\circ}$.  The total
driving force (black), equilibrium (blue), non-equilibrium (green), and volumetric (red)
components are plotted separately to isolate each physical contribution.

For the balanced distribution (Fig.~\ref{fig:driving_force0}), the four-lobed polar
pattern reflects the quasi-isotropic symmetry of the $\pm 45^{\circ}$ architecture, with
maximum energy accumulation along the $0^{\circ}$ and $90^{\circ}$ directions.  As viscous
relaxation proceeds from $t = 10^{-6}$\,s (Fig.~\ref{fig:driving_force0}a) to
near-equilibrium at $t = 0.1$\,s (Fig.~\ref{fig:driving_force0}d), the total driving
force undergoes approximately 40\% reduction, driven primarily by the decay of the
non-equilibrium (viscous) contribution.  The equilibrium component remains nearly constant
throughout, while the volumetric contribution grows monotonically to become the dominant
term at later times, reflecting the progressive build-up of tensile hydrostatic stress.
For the unbalanced distribution (Fig.~\ref{fig:driving_force}), the directional bias
imposed by the dominant $-45^{\circ}$ family breaks the fourfold symmetry, producing an
asymmetric pattern in which the lobe aligned with the $-45^{\circ}$ fiber direction is
noticeably more elongated.  Crucially, this directional anisotropy is preserved throughout
the entire relaxation history (Figs.~\ref{fig:driving_force}a--d), demonstrating that
fiber architecture fundamentally controls the spatial distribution of the crack driving
force irrespective of the viscous state of the matrix.

\begin{table}[H]
	\centering
	\caption{Material parameters used in the numerical simulations.}
	\label{tab:param}
	\begin{footnotesize}
		\begin{tabular}{lcccc}
			\toprule
			Parameter & Symbol & Value & Equation & References \\
			\midrule
			\multicolumn{5}{l}{\textit{Viscoelastic--viscoplastic  model}} \\
			\midrule
			Equilibrium shear modulus      & $\mu_{\text{eq}}^{0}$ (MPa)      & 760                   & (\ref{eq:psi_matrix})  & \cite{bahtiri2024thermodynamically} \\
			Non-equilibrium shear modulus  & $\mu_{\text{neq}}^{0}$ (MPa)     & 790                   & (\ref{eq:psi_matrix})  & \cite{bahtiri2024thermodynamically} \\
			Volumetric bulk modulus        & $k_{v}^{0}$ (MPa)                & 1154                  & (\ref{eq:psi_vol})     & \cite{bahtiri2024thermodynamically} \\
			Viscoelastic pre-factor        & $\dot{\varepsilon}_{0}$ (s$^{-1}$) & $1.0447\times10^{12}$ & (\ref{eq:eps_dot_v})  & \cite{bahtiri2024thermodynamically} \\
			Activation energy              & $\Delta H$ (J)                   & $1.977\times10^{-19}$ & (\ref{eq:eps_dot_v})   & \cite{bahtiri2024thermodynamically} \\
			Stress exponent                & $m$                              & 0.657                 & (\ref{eq:eps_dot_v})   & \cite{bahtiri2024thermodynamically} \\
			Athermal yield stress          & $\tau_{0}$ (MPa)                 & 40                    & (\ref{eq:eps_dot_v})   & \cite{bahtiri2024thermodynamically} \\
			Viscoplastic parameter         & $a$                              & 0.22$\omega_w$ + 0.005  & (\ref{eq:eps_dot_vp})  & --                       \\
			Viscoplastic exponent          & $b$                              & 1.1                   & (\ref{eq:eps_dot_vp})  & --                       \\
			Viscoplastic threshold         & $\sigma_{0}$ (MPa)               & \cite{bahtiri2024thermodynamically}                    & (\ref{eq:eps_dot_vp})  & --                       \\
			Temperature sensitivity        & $\alpha_{\theta}$ (K$^{-1}$)              & 0.01093               & (\ref{eq:J_theta}) & \cite{bahtiri2024thermodynamically} \\
			Fiber stiffness parameter      & $a_1$                            & 9                     & (\ref{eq:f_function})  & \cite{arash2019viscoelastic2}                       \\
			Fiber stiffness parameter      & $a_2$                            & 1                     & (\ref{eq:f_function})  & \cite{arash2019viscoelastic2}                       \\
			Fiber stiffness parameter      & $a_3$                            & 1                     & (\ref{eq:f_function})  & \cite{arash2019viscoelastic2}                       \\
			\midrule
			\multicolumn{5}{l}{\textit{Phase-field fracture model}} \\
			\midrule
			Critical energy release rate   & $G_{c}$ (N/mm)                   & 0.200                 & (\ref{eq:phasefield_spat}) & \cite{arash2023effect} \\
			Length scale parameter         & $l_{0}$ (mm)                     & 0.02                  & (\ref{eq:phasefield_spat}) & \cite{arash2023effect} \\
			\midrule
			\multicolumn{5}{l}{\textit{Piezoresistive electrical model}} \\
			\midrule
			Matrix conductivity            & $\sigma_m$ (S/mm)                & $10^{-14}$            & (\ref{eq:sigma0_tensor})      & --                       \\
			Axial fiber conductivity       & $\sigma^0_{\parallel}$ (S/mm)    & 66.7                  & (\ref{eq:sigma_par})   & \cite{ji2022anisotropic}            \\
			Transverse fiber conductivity  & $\sigma^0_{\perp}$ (S/mm)       & 15.9                  & (\ref{eq:sigma_perp})  & \cite{ji2022anisotropic}            \\
			Longitudinal gauge factor      & $\text{GF}_{\parallel}$          & 2.0                   & (\ref{eq:sigma_par})   & \cite{abry1999situ}          \\
			Transverse gauge factor        & $\text{GF}_{\perp}$              & 2.0                   & (\ref{eq:sigma_perp})  & \cite{abry1999situ}          \\
			Electrical degradation exponent & $p$                             & 2                     & (\ref{eq:h_elec})          & --                       \\
			Residual conductivity          & $k_e$                            & $10^{-6}$             & (\ref{eq:h_elec})          & --                       \\
			Applied voltage                & $V_{\text{app}}$ (V)             & 1.0                   & (\ref{eq:dbc_elec})     & --                       \\
			\bottomrule
		\end{tabular}
	\end{footnotesize}
\end{table}


\begin{figure}[H]
\centering
	\scalebox{0.7}{%
		\includegraphics{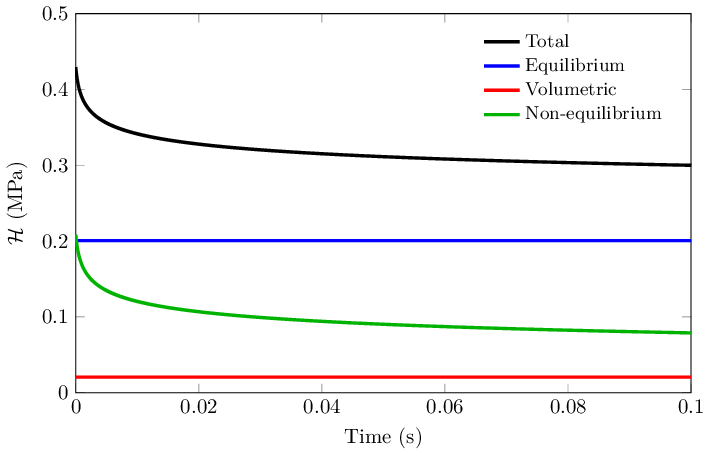}
	}
\caption{Temporal evolution of crack driving force components for CF (50 wt\%)/epoxy with unbalanced fiber distribution (70 wt\% at $-45^{\circ}$ and 30 wt\% at $+45^{\circ}$) at 296~K under uniaxial tensile loading applied at $45^{\circ}$ to the horizontal axis.}
\label{fig:energy_evolution}
\end{figure}

The temporal evolution of the individual energy contributions for the unbalanced
configuration is summarized in Fig.~\ref{fig:energy_evolution}.  The total driving force
$\mathcal{H}$ rises steeply upon loading and then decays as the non-equilibrium (viscous)
energy relaxes, decreasing from its initial peak by approximately 55\% at $t = 0.1$\,s.
The equilibrium energy reaches a constant value immediately after loading and remains
unchanged throughout relaxation, confirming that the equilibrium network stores energy
elastically without dissipation.  The volumetric energy—initially negligible—grows
monotonically to become the dominant contributor at long times, indicating that the
transition from rate-dependent to quasi-elastic fracture behavior is governed by
hydrostatic tension build-up.  This decomposition confirms a fundamental feature of
fracture in viscoelastic-viscoplastic composites: the crack driving force is the sum of
time-varying contributions with distinct physical origins, a coupling that cannot be
captured by elastic phase-field models.

\begin{figure}[H]
	\centering
	%
	\begin{subfigure}[b]{0.48\linewidth}
		\centering
		\includegraphics{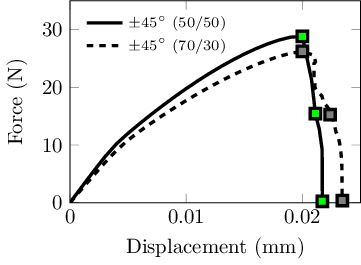}
		\caption{}
		\label{fig:fd_conductance_45_1}
	\end{subfigure}%
	%
	\begin{subfigure}[b]{0.48\linewidth}
		\centering
		\includegraphics{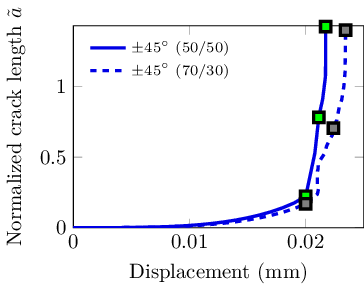}
		\caption{}
		\label{fig:fd_conductance_45_2}
	\end{subfigure}
	%
	\begin{subfigure}[b]{0.48\linewidth}
		\centering
		\includegraphics{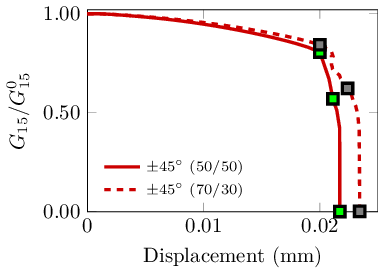}
		\caption{}
		\label{fig:fd_conductance_45_3}
	\end{subfigure}
	%
	\begin{subfigure}[b]{0.48\linewidth}
		\centering
		\includegraphics{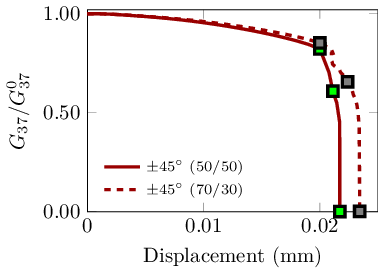}
		\caption{}
		\label{fig:fd_conductance_45_4}
	\end{subfigure}%
	\caption{Force--displacement response, normalized crack length
		$\tilde{a}$, normalized compliance $\tilde{C}$, and
		normalized conductances $G_{ij}/G_{ij}^{0}$
		vs.\ applied displacement for CF~(30~wt\%)/epoxy composite at 298~K.}
	\label{fig:fd_conductance_45}
\end{figure}

Figure~\ref{fig:fd_conductance_45} presents the electromechanical response
of CF~(30~wt\%)/epoxy composites with two $\pm45^\circ$ fiber
configurations, balanced (50/50) and unbalanced (70/30) weight
distributions, at 298~K.
The force--displacement curves in Fig.~\ref{fig:fd_conductance_45_1}
show that both configurations exhibit similar peak loads and initial
stiffness, reflecting the equal total fiber content. The slight softening
of the 70/30 case arises from the asymmetric crack resistance introduced
by the dominant $-45^\circ$ family, which reduces the effective fracture
toughness in the loading direction.
The normalized crack length $\tilde{a}$ in
Fig.~\ref{fig:fd_conductance_45_2} remains near zero throughout the
pre-peak regime and increases sharply at the onset of unstable crack
propagation, confirming that crack growth is concentrated in a narrow
displacement interval.
The 70/30 configuration fractures at a slightly lower displacement than
the balanced case, consistent with the reduced resistance offered by the
dominant fiber family to crack advance along its preferred direction.

The normalized conductances $G_{15}/G_{15}^0$ and $G_{37}/G_{37}^0$ in
Figs.~\ref{fig:fd_conductance_45_3}--\ref{fig:fd_conductance_45_4} decrease monotonically from
unity throughout loading, reflecting the progressive degradation of the
effective conductivity tensor $\hat{\boldsymbol{\sigma}}_\mathrm{eff}$
driven by the accumulation of $\phi$.
In the pre-fracture regime both configurations produce nearly identical
conductance trajectories, indicating that the gradual pre-peak damage
accumulation is insensitive to the weight ratio asymmetry. The 70/30 configuration produces a steeper conductance drop at an earlier
displacement, encoding the earlier fracture displacement directly in the
electrical response.
Crucially, the two electrode pairs $G_{15}/G_{15}^0$ and
$G_{37}/G_{37}^0$, respectively oriented along the main diagonal and horizontally, respond differently to the two configurations. The
horizontal pair $G_{37}/G_{37}^0$ is more sensitive to the asymmetric
crack path driven by the dominant $-45^\circ$ family, while the diagonal
pair $G_{15}/G_{15}^0$ responds more symmetrically.
This directional sensitivity, replicated across all 28 electrode pairs,
provides the basis for the inverse analysis described in
Section~\ref{sec:shm_inverse}. Different fiber architectures produce
distinguishable conductance signatures, and the multi-electrode network
resolves orientation-dependent damage features that a single electrode
pair cannot capture.

\begin{figure}[H]
	\centering
	\begin{subfigure}[b]{1\linewidth}
		\centering
		\includegraphics{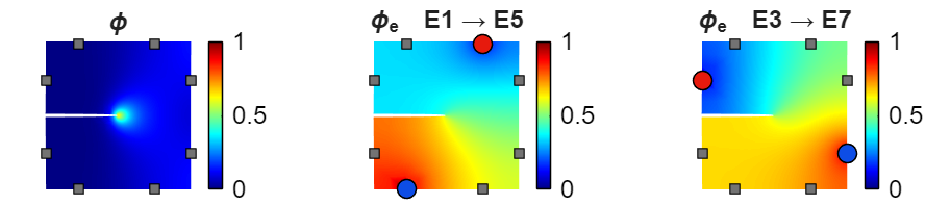}
		\caption{ }
		\label{fig:fracture_CF03_-45_45_50-50_1}
	\end{subfigure}
	
	\begin{subfigure}[b]{1\linewidth}
		\centering
		\includegraphics{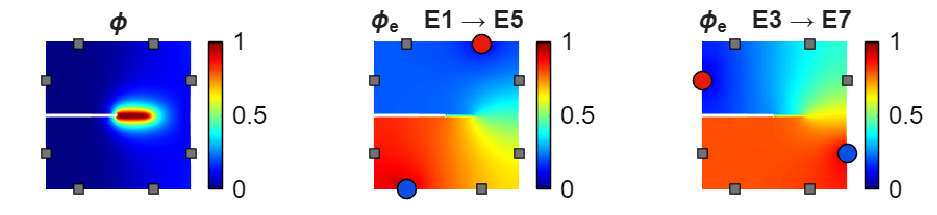}
		\caption{ }
		\label{fig:fracture_CF03_-45_45_50-50_2}
	\end{subfigure}
	
	\begin{subfigure}[b]{1\linewidth}
		\centering	
		\includegraphics{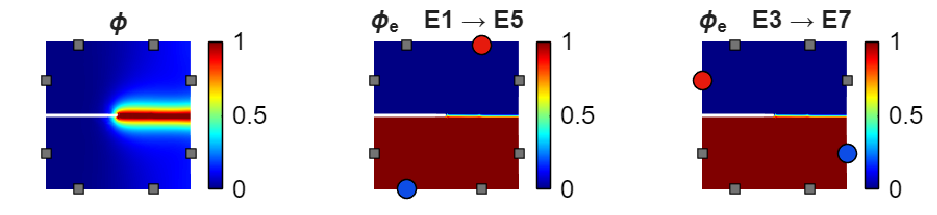}
		\caption{}			
		\label{fig:fracture_CF03_-45_45_50-50_3}
	\end{subfigure}		
	\caption{Crack propagation in CF~(30~wt\%)/epoxy composite
		with $\pm45^{\circ}$ (50/50) fibre configuration at displacements: (a) 0.0200~mm, (b) 0.0211~mm, and (c) 0.0217~mm. The snapshots are marked at the force-displacement response in Fig.~6.}
	\label{fig:contours_4550}
\end{figure}

\begin{figure}[H]
	\centering
	\begin{subfigure}[b]{1\linewidth}
		\centering
		\includegraphics{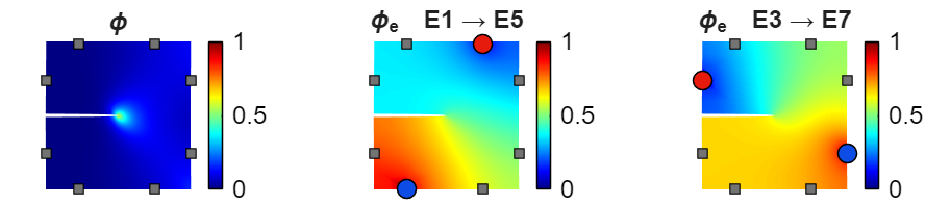}
		\caption{ }
		\label{fig:fracture_CF03_-45_45_70-30_1}
	\end{subfigure}
	
	\begin{subfigure}[b]{1\linewidth}
		\centering
		\includegraphics{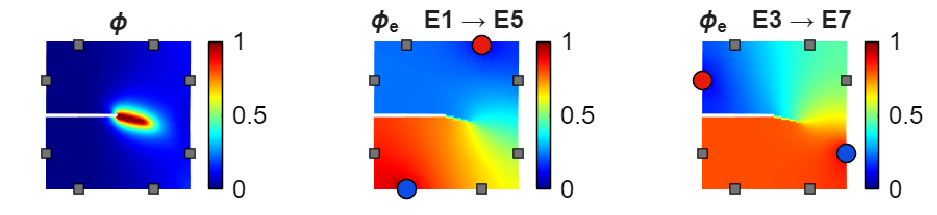}
		\caption{ }
		\label{fig:fracture_CF03_-45_45_70-30_2}
	\end{subfigure}
	
	\begin{subfigure}[b]{1\linewidth}
		\centering	
		\includegraphics{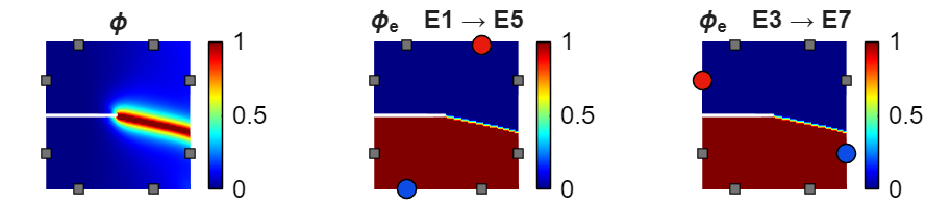}
		\caption{}			
		\label{fig:fracture_CF03_-45_45_70-30_3}
	\end{subfigure}		
	\caption{Crack propagation in CF~(30~wt\%)/epoxy composite
		with $\pm45^{\circ}$ (70/30) fibre configuration at displacements: (a) 0.0200~mm, (b) 0.0224~mm, and (c) 0.0234~mm. The snapshots are marked at the force-displacement response in Fig.~6.}
	\label{fig:contours_4570}
\end{figure}

The spatial evolution of the phase-field variable and electric potential
fields at the three marked displacements is shown in
Figs.~\ref{fig:contours_4550} and~\ref{fig:contours_4570} for the
balanced and unbalanced configurations, respectively.
In both cases the phase-field contours confirm that crack propagation
is confined to a narrow band oriented along the dominant fibre family:
the balanced 50/50 configuration produces a nominally horizontal crack
path, while the dominant $-45^\circ$ family in the 70/30 case deflects
the crack downward at approximately $-45^\circ$ to the horizontal.
The electric potential fields $\phi_e$ for the E1$\to$E5 (main diagonal)
and E3$\to$E7 (horizontal) pairs respond distinctly to this difference:
the horizontal pair is strongly disrupted by the oblique crack in the
70/30 case, producing a more pronounced potential discontinuity across
the crack plane, whereas the diagonal pair is comparatively less affected.
This orientation-dependent distortion of the potential field is the
physical mechanism by which the 28-pair conductance network encodes
crack path information, and it motivates the use of the full electrode
set as input to the inverse model.

\begin{figure}[H]
	\centering
	%
	\begin{subfigure}[b]{0.48\linewidth}
		\centering
		\includegraphics{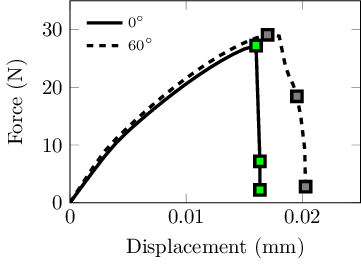}
		\caption{}
		\label{fig:fd_conductance_0_60_1}
	\end{subfigure}%
	%
	\begin{subfigure}[b]{0.48\linewidth}
		\centering
		\includegraphics{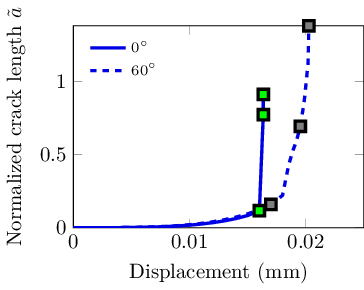}
		\caption{}
		\label{fig:fd_conductance_0_60_2}
	\end{subfigure}
	%
	\begin{subfigure}[b]{0.48\linewidth}
		\centering
		\includegraphics{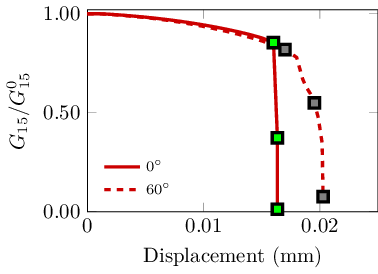}
		\caption{}
		\label{fig:fd_conductance_0_60_3}
	\end{subfigure}
	%
	\begin{subfigure}[b]{0.48\linewidth}
		\centering
		\includegraphics{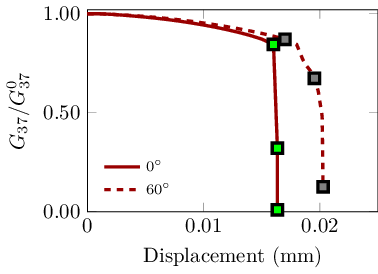}
		\caption{}
		\label{fig:fd_conductance_0_60_4}
	\end{subfigure}%
	\caption{Force--displacement response, normalized crack length
		$\tilde{a}$, normalized compliance $\tilde{C}$, and
		normalized conductances $G_{ij}/G_{ij}^{0}$
		vs.\ applied displacement for CF~(30~wt\%)/epoxy composite at 298~K.}
	\label{fig:fd_conductance_0_60}
\end{figure}

Figure~\ref{fig:fd_conductance_0_60} presents the electromechanical
response of CF~(30~wt\%)/epoxy composites with single-fiber-family
configurations aligned at $0^\circ$ and $60^\circ$ at 298~K.
The force--displacement curves in Fig.~\ref{fig:fd_conductance_0_60_1}
show that both configurations exhibit similar overall stiffness and peak
load, reflecting the identical fiber content and the dominance of the
matrix response in the pre-peak regime.
The $0^\circ$ configuration fractures at a slightly lower displacement
than the $60^\circ$ case, as the horizontal fiber alignment offers
minimal resistance to mode-I crack opening perpendicular to the loading
direction, while the $60^\circ$ fibers provide a partial deflection
mechanism that marginally delays unstable crack propagation.

The normalized crack length $\tilde{a}$ in
Fig.~\ref{fig:fd_conductance_0_60}(b) exhibits the same characteristic
pattern observed in Fig.~\ref{fig:fd_conductance_45}: negligible growth
throughout the pre-peak regime followed by a sharp jump at fracture.
However, the post-fracture $\tilde{a}$ values differ between the two
configurations, reflecting the different crack path lengths associated
with a horizontal crack ($0^\circ$) and an oblique crack ($60^\circ$)
traversing the specimen width.

\begin{figure}[H]
	\centering
	\begin{subfigure}[b]{1\linewidth}
		\centering
		\includegraphics{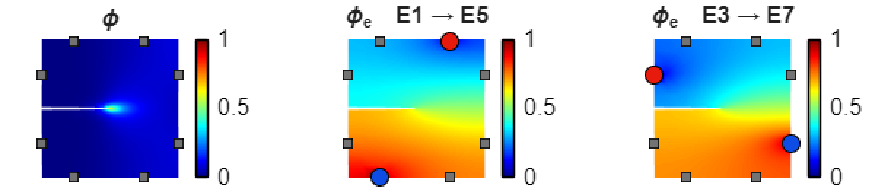}
		\caption{ }
		\label{fig:fracture_CF03_0_1}
	\end{subfigure}
	
	\begin{subfigure}[b]{1\linewidth}
		\centering
		\includegraphics{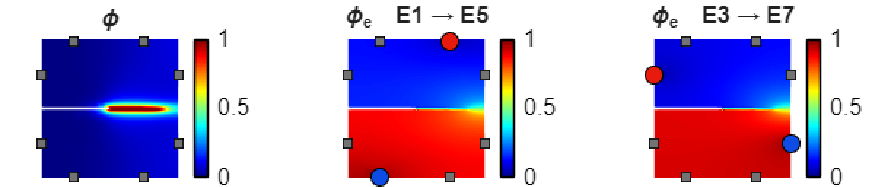}
		\caption{ }
		\label{fig:fracture_CF03_0_2}
	\end{subfigure}
	
	\begin{subfigure}[b]{1\linewidth}
		\centering	
		\includegraphics{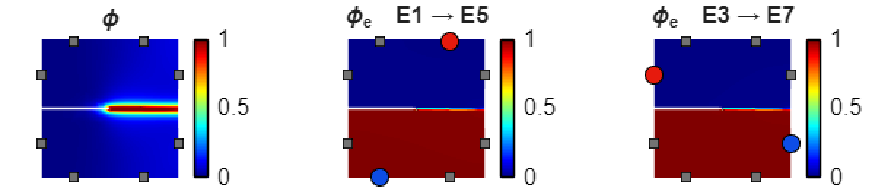}
		\caption{}			
		\label{fig:fracture_CF03_0_3}
	\end{subfigure}		
	\caption{Crack propagation in CF~(30~wt\%)/epoxy composite
		with $0^{\circ}$ fibre configuration at displacements: (a) 0.0160~mm, (b) 0.0163~mm, and (c) 0.0164~mm. The snapshots are marked at the force-displacement response in Fig.~9.}
	\label{fig:contours_0deg}
\end{figure}

\begin{figure}[H]
	\centering
	\begin{subfigure}[b]{1\linewidth}
		\centering
		\includegraphics{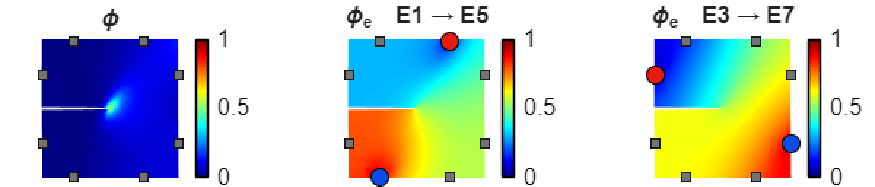}
		\caption{ }
		\label{fig:fracture_CF03_60_1}
	\end{subfigure}
	
	\begin{subfigure}[b]{1\linewidth}
		\centering
		\includegraphics{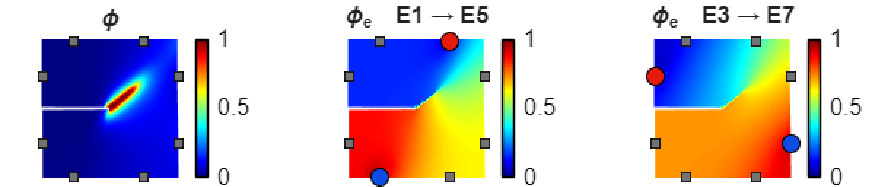}
		\caption{ }
		\label{fig:fracture_CF03_60_2}
	\end{subfigure}
	
	\begin{subfigure}[b]{1\linewidth}
		\centering	
		\includegraphics{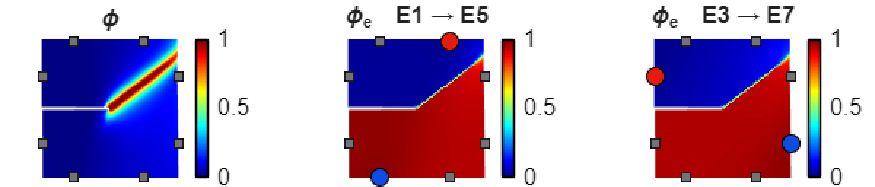}
		\caption{}			
		\label{fig:fracture_CF03_60_3}
	\end{subfigure}		
	\caption{Crack propagation in CF~(30~wt\%)/epoxy composite
		with $60^{\circ}$ ibre configuration at displacements: (a) 0.0200~mm, (b) 0.0224~mm, and (c) 0.0234~mm. The snapshots are marked at the force-displacement response in Fig.~9.}
	\label{fig:contours_60deg}
\end{figure}

The normalized conductances $G_{15}/G_{15}^0$ and $G_{37}/G_{37}^0$ in
Figs.~\ref{fig:fd_conductance_0_60}(c)--(d) reveal a qualitatively
different behavior from the $\pm45^\circ$ cases.
In the pre-fracture regime both configurations again produce similar
trajectories, but at fracture the two pairs respond in opposite fashion
depending on the crack orientation.
For the $0^\circ$ configuration, the horizontal crack propagates
perpendicular to the E1$\to$E5 diagonal pair and parallel to the
E3$\to$E7 horizontal pair: the diagonal pair therefore experiences a
complete severing of the current path, producing an abrupt and near-total
conductance collapse, while the horizontal pair is comparatively less
disrupted.
For the $60^\circ$ configuration the crack deflects toward the fiber
direction, cutting obliquely across both pairs and producing intermediate
conductance drops in both, with the diagonal pair now less affected than
in the $0^\circ$ case.
This contrast between the two configurations — barely distinguishable in
the force--displacement response but clearly separated in the directional
conductance signatures — directly demonstrates the orientation sensitivity
of the multi-electrode network and the insufficiency of mechanical
measurements alone for fiber architecture identification.
The three displacement snapshots marked in
Fig.~\ref{fig:fd_conductance_0_60} correspond to the phase-field and
electric potential contours shown in
Figs.~\ref{fig:contours_0deg} and~\ref{fig:contours_60deg}, which confirm
the distinct spatial crack paths and their contrasting effects on the
electric potential distributions for the two electrode pairs shown.

\pgfplotsset{
	eit axis/.style={
		enlargelimits    = false,
		unbounded coords = discard,
		xmin          = 0,
		xmax          = 0.025,
		xtick         = {0, 0.01, 0.02},
		scaled ticks  = false,
		xticklabels   = {0, 0.01, 0.02},
		xlabel        = {Displacement (mm)},
		width         = 6.5cm,
		height        = 5.0cm,
		legend pos    = north east,
		legend style  = {draw=none, fill=none, font=\small},
		legend cell align = {left},
	},
	eit conductance axis/.style={
		eit axis,
		ymin  = 0,
		ymax  = 1.02,
		ytick = {0, 0.5, 1.0},
		yticklabel style = {
			/pgf/number format/fixed,
			/pgf/number format/fixed zerofill,
			/pgf/number format/precision=2,
		},
	},
	eit line/.style={
		solid,
		mark       = none,
		line width = 1.5pt,
	},
	eit markers/.style={
		only marks,
		mark         = square*,
		mark size    = 2.5pt,
		mark options = {draw=black, fill=green, line width=1.5pt},
		x filter/.code={
			\edef\mystep{\thisrowno{0}}%
			\pgfmathparse{
				(abs(\mystep - 1600) < 0.5) ||
				(abs(\mystep - 1679) < 0.5) ||
				(abs(\mystep - 1684) < 0.5)
				? \pgfmathresult : nan}%
		},
	},
	eit markers2/.style={
		only marks,
		mark         = square*,
		mark size    = 2.5pt,
		mark options = {draw=black, fill=gray, line width=1.5pt},
		x filter/.code={
			\edef\mystep{\thisrowno{0}}%
			\pgfmathparse{
				(abs(\mystep - 1700) < 0.5) ||
				(abs(\mystep - 2200) < 0.5) ||
				(abs(\mystep - 2430) < 0.5)
				? \pgfmathresult : nan}%
		},
	},
}

\begin{figure}[H]
	\centering
	%
	\begin{subfigure}[b]{0.48\linewidth}
		\centering
		\includegraphics{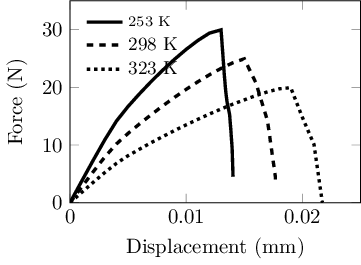}
		\caption{}
		\label{fig:fd_conductance_temp_1}
	\end{subfigure}%
	%
	\begin{subfigure}[b]{0.48\linewidth}
		\centering
		\includegraphics{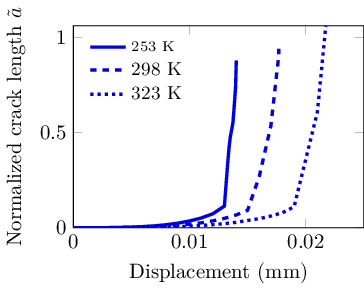}
		\caption{}
		\label{fig:fd_conductance_temp_2}
	\end{subfigure}
	%
	\begin{subfigure}[b]{0.48\linewidth}
		\centering
		\includegraphics{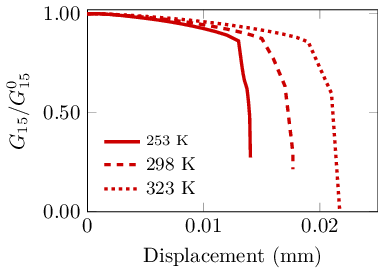}
		\caption{}
		\label{fig:fd_conductance_temp_3}
	\end{subfigure}
	%
	\begin{subfigure}[b]{0.48\linewidth}
		\centering
		\includegraphics{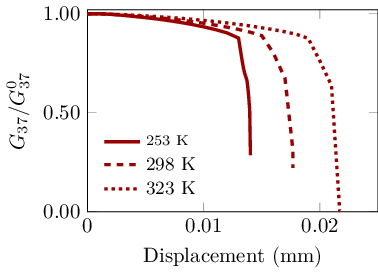}
		\caption{}
		\label{fig:fd_conductance_temp_4}
	\end{subfigure}%
	\caption{Effect of temperature on the force--displacement response, normalized crack length
		$\tilde{a}$, normalized compliance $\tilde{C}$, and
		normalized conductances $G_{ij}/G_{ij}^{0}$
		vs.\ applied displacement for CF~(30~wt\%)/epoxy composite.}
	\label{fig:fd_conductance_temp}
\end{figure}

Figure~\ref{fig:fd_conductance_temp} examines the effect of temperature
on the electromechanical response of CF~(30~wt\%)/epoxy composites with
randomly oriented fibres at $v_\mathrm{CF} = 0.3$, spanning the
temperature range $\theta \in \{253, 298, 323\}$~K relevant to structural
service conditions.

The force--displacement curves in
Fig.~\ref{fig:fd_conductance_temp_1} show a pronounced temperature
dependence of both stiffness and peak load: lowering the temperature from
323~K to 253~K increases the peak force from approximately 20~N to 30~N
and reduces the fracture displacement, reflecting the stiffening and
embrittlement of the epoxy matrix at sub-ambient temperatures governed
by the Arrhenius-type viscoelastic flow rule
(\eqref{eq:eps_dot_v}) and the temperature-sensitive viscoplastic
threshold.
The normalized crack length $\tilde{a}$ in
Fig.~\ref{fig:fd_conductance_temp_2} exhibits a consistent sharp-onset
pattern across all three temperatures, with the fracture displacement
shifting to higher values as temperature increases in accordance with
the thermally governed failure displacement identified in the preceding
sections.

The normalized conductances $G_{15}/G_{15}^0$ and $G_{37}/G_{37}^0$
in Figs.~\ref{fig:fd_conductance_temp_3}--\ref{fig:fd_conductance_temp_4} reveal that the pre-fracture conductance trajectories are
nearly identical across all three temperatures when plotted against
displacement. However, the displacement at which the abrupt conductance drop occurs shifts with temperature. Consequently, when the conductance ratios are mapped against the normalized crack length $\tilde{a}$ rather than displacement, the three curves collapse onto a single trajectory — confirming that
$\tilde{a}$ is temperature-invariant as a function of the electrical
observables. The normalized compliance $\tilde{C}$, by contrast, evolves at a
different rate with displacement at each temperature because the
stiffness degradation rate is governed by the temperature-dependent
viscoelastic--viscoplastic material response, and this rate difference
is not captured by the conductance ratios alone.
This distinction motivates including temperature $T$ as an explicit
input to the inverse model alongside the fiber architecture descriptors
and the conductance measurements, as described in
Section~\ref{sec:ann_training}.

\pgfplotsset{
	eit axis/.style={
		enlargelimits    = false,
		unbounded coords = discard,
		xmin          = 0,
		xmax          = 0.025,
		xtick         = {0, 0.01, 0.02},
		scaled ticks  = false,
		xticklabels   = {0, 0.01, 0.02},
		xlabel        = {Displacement (mm)},
		width         = 6.5cm,
		height        = 5.0cm,
		legend pos    = north east,
		legend style  = {draw=none, fill=none, font=\small},
		legend cell align = {left},
	},
	eit conductance axis/.style={
		eit axis,
		ymin  = 0,
		ymax  = 1.02,
		ytick = {0, 0.5, 1.0},
		yticklabel style = {
			/pgf/number format/fixed,
			/pgf/number format/fixed zerofill,
			/pgf/number format/precision=2,
		},
	},
	eit line/.style={
		solid,
		mark       = none,
		line width = 1.5pt,
	},
	eit markers/.style={
		only marks,
		mark         = square*,
		mark size    = 2.5pt,
		mark options = {draw=black, fill=green, line width=1.5pt},
		x filter/.code={
			\edef\mystep{\thisrowno{0}}%
			\pgfmathparse{
				(abs(\mystep - 1600) < 0.5) ||
				(abs(\mystep - 1679) < 0.5) ||
				(abs(\mystep - 1684) < 0.5)
				? \pgfmathresult : nan}%
		},
	},
	eit markers2/.style={
		only marks,
		mark         = square*,
		mark size    = 2.5pt,
		mark options = {draw=black, fill=gray, line width=1.5pt},
		x filter/.code={
			\edef\mystep{\thisrowno{0}}%
			\pgfmathparse{
				(abs(\mystep - 1700) < 0.5) ||
				(abs(\mystep - 2200) < 0.5) ||
				(abs(\mystep - 2430) < 0.5)
				? \pgfmathresult : nan}%
		},
	},
}

\subsection{Dataset assembly and ANN model training}
\label{sec:ann_training}

The training dataset is assembled from a systematic set of phase-field
simulations on the single-edge notched specimen described in
Section~\ref{sec:shm_inverse}, covering a broad range of fiber
architectures, carbon fiber contents, and temperatures representative of
injection-moulded SCFRP composites in service conditions.
All simulations are performed at three temperatures
$T \in \{253, 298, 323\}$~K to capture the effect of thermal loading on
the damage-induced piezoresistive response.
The simulated configurations are summarised in Table~\ref{tab:dataset}.
Single-fiber-family cases are generated for eight orientations
$\Theta \in \{-60^\circ, -45^\circ, -30^\circ, 0^\circ,
30^\circ, 45^\circ, 60^\circ, 90^\circ\}$ at $v_\mathrm{CF} = 0.3$.
Two-fiber-family cases with $\pm 45^\circ$ and $0^\circ/90^\circ$
are generated for three weight ratios (30/70, 50/50, 70/30), and an
additional two-family case covers the $0^\circ/60^\circ$ (50/50)
configuration, all at $v_\mathrm{CF} = 0.3$.
Randomly oriented fibers are simulated at three fiber contents
$v_\mathrm{CF} \in \{0.1, 0.3, 0.5\}$.
Two cases are held out entirely from training and used exclusively for
validation: the $0^\circ/60^\circ$ (50/50) two-family configuration at
298~K, and randomly oriented fibers at $v_\mathrm{CF} = 0.3$ across all
three temperatures.

\begin{table}[H]
	\centering
	\caption{Summary of phase-field simulation cases used for dataset
		assembly. Test cases are excluded from training.}
	\label{tab:dataset}
	\small
	\begin{tabular}{llcc}
		\hline
		Fiber architecture & $v_\mathrm{CF}$ & $\theta$ (K) & Role \\
		\hline
		Single family, $\Theta = -60^\circ, \ldots, 90^\circ$ (8 cases)
		& 0.3 & 253, 298, 323 & Training \\
		Two family, $\pm 45^\circ$ (30/70, 50/50, 70/30)
		& 0.3 & 253, 298, 323 & Training \\
		Two family, $0^\circ/90^\circ$ (30/70, 50/50, 70/30)
		& 0.3 & 253, 298, 323 & Training \\
		Two family, $0^\circ/60^\circ$ (50/50)
		& 0.3 & 253, 323      & Training \\
		Two family, $0^\circ/60^\circ$ (50/50)
		& 0.3 & 298           & \textbf{Test} \\
		Random orientation
		& 0.1, 0.5 & 253, 298, 323 & Training \\
		Random orientation
		& 0.3      & 253, 323      & Training \\
		Random orientation
		& 0.3      & 298           & \textbf{Test} \\
		\hline
	\end{tabular}
\end{table}

For each simulation case, post-processing yields one row per saved load
step containing the inputs and outputs defined in
Eq.~(\ref{eq:inverse_map}).
Only load steps satisfying $x_\mathrm{tip}/W < 1$ are retained,
discarding fully fractured states in which the crack has reached the far
boundary and the compliance $\tilde{C}$ diverges numerically.
The initial undamaged state (step~0) is included as a reference row with
$\tilde{a} = 0$, $\tilde{C} = 1$, and
$G_{ij}^{(0)}/G_{ij}^0 = 1$ for all pairs,
providing the model with an explicit representation of the unloaded
reference.
The assembled training set contains $N_\mathrm{train}$ rows after
filtering, and the test set contains $N_\mathrm{test}$ rows from the
held-out configuration.

A key design choice is the representation of fiber architecture in the input vector.
A scalar fiber angle $\theta$ is insufficient for two-family configurations with
unequal weight ratios: the pairs $(-45^\circ, 70\%) + (45^\circ, 30\%)$ and
$(-45^\circ, 30\%) + (45^\circ, 70\%)$ are indistinguishable by $\theta$ alone.
The second-order orientation tensor $\mathbf{A}$ resolves
this ambiguity exactly, as it encodes both the principal fiber directions and their
relative weights through its eigenvalue decomposition.
For the 2D in-plane problem, $\mathbf{A}$ has two independent components ($A_{11}$
and $A_{12}$, with $A_{11} + A_{22} = 1$), which serve as the microstructure
descriptors in the input vector.
This parameterisation is consistent with the role of $\mathbf{A}$ in the forward
model, where the same tensor simultaneously defines the anisotropic crack resistance
(Section~\ref{sec:phase}) and the principal conduction directions of the carbon fiber
network (Section~\ref{sec:elec}).
The orientation tensor components for all simulated configurations are listed in
Table~\ref{tab:A_values}.

It should be noted that several distinct fiber architectures map to the same second-order orientation tensor components. Specifically, the
$\pm45^\circ$ (50/50), $0^\circ/90^\circ$ (50/50), and randomly oriented configurations all yield $(A_{11}, A_{12}) = (0.500, 0.000)$, and are therefore indistinguishable at the level of $\mathbf{A}$. Higher-order orientation tensors would resolve these configurations, but at the cost of a significantly larger input dimension and the need for substantially more training data. Within the present framework this degeneracy is acceptable because the three configurations also produce similar conductance signatures. This degeneracy is acknowledged as a limitation of the second-order tensor parameterisation and motivates future work incorporating higher-order microstructure descriptors.

\begin{table}[H]
	\centering
	\caption{Orientation tensor components $(A_{11}, A_{12})$ for all
		simulated fiber configurations.}
	\label{tab:A_values}
	\begin{tabular}{lcc}
		\hline
		Configuration & $A_{11}$ & $A_{12}$ \\
		\hline
		Single family, $-60^\circ$             & 0.250 & $-0.433$ \\
		Single family, $-45^\circ$             & 0.500 & $-0.500$ \\
		Single family, $-30^\circ$             & 0.750 & $-0.433$ \\
		Single family, $0^\circ$               & 1.000 &  0.000   \\
		Single family, $30^\circ$              & 0.750 &  0.433   \\
		Single family, $45^\circ$              & 0.500 &  0.500   \\
		Single family, $60^\circ$              & 0.250 &  0.433   \\
		Single family, $90^\circ$              & 0.000 &  0.000   \\
		Two family, $\pm45^\circ$ (30/70)      & 0.500 &  0.200   \\
		Two family, $\pm45^\circ$ (50/50)      & 0.500 &  0.000   \\
		Two family, $\pm45^\circ$ (70/30)      & 0.500 & $-0.200$ \\
		Two family, $0^\circ/90^\circ$ (30/70) & 0.300 &  0.000   \\
		Two family, $0^\circ/90^\circ$ (50/50) & 0.500 &  0.000   \\
		Two family, $0^\circ/90^\circ$ (70/30) & 0.700 &  0.000   \\
		Two family, $0^\circ/60^\circ$ (50/50) & 0.625 &  0.217   \\
		Random (isotropic)                     & 0.500 &  0.000   \\
		\hline
	\end{tabular}
\end{table}

The inverse mapping $\mathcal{F}$ (Eq.~(\ref{eq:inverse_map})) is
approximated by a feedforward ANN with the following structure.
The input layer receives the 31-dimensional vector
\begin{equation}
	\mathbf{x}^{(n)} =
	\bigl(
	A_{11},\; A_{12},\; v_\mathrm{CF},\;
	G_{12}^{(n)}/G_{12}^0,\; \ldots,\; G_{78}^{(n)}/G_{78}^0
	\bigr)
	\in \mathbb{R}^{31},
	\label{eq:ann_input}
\end{equation}
comprising three microstructure descriptors and 28 normalised
conductance ratios for the $\binom{8}{2} = 28$ electrode pairs.
The output layer produces the two-dimensional vector
$\hat{\mathbf{y}}^{(n)} = (\hat{\tilde{a}}^{(n)},\;
\hat{\tilde{C}}^{(n)}) \in \mathbb{R}^2$.
Two fully connected hidden layers with 16 neurons each and hyperbolic
tangent activation functions connect input to output; the output layer
uses a linear activation.
The total number of trainable parameters is
$(31 \times 16 + 16) + (16 \times 16 + 16) + (16 \times 2 + 2) = 786$.

All inputs and outputs are z-score normalised prior to training using
the mean and standard deviation of the training set exclusively,
preventing any leakage of test-set statistics into the model.
The network is trained by minimising the mean squared error (MSE)
between predicted and true normalised outputs using the
Levenberg--Marquardt algorithm~\cite{levenberg1944method}, which approximates the Hessian of the loss through the Gauss--Newton approximation
$\mathbf{H} \approx \mathbf{J}^\top \mathbf{J}$ and is well-suited for
small-to-medium regression networks where the full Jacobian can be
formed.
Training is terminated by early stopping with a patience of 20 epochs,
monitoring the MSE on a 15\% validation split drawn randomly from the
training set.
The maximum number of training epochs is set to 1000. Convergence is
consistently achieved before this limit.

\begin{figure}[H]
	\centering
	\includegraphics{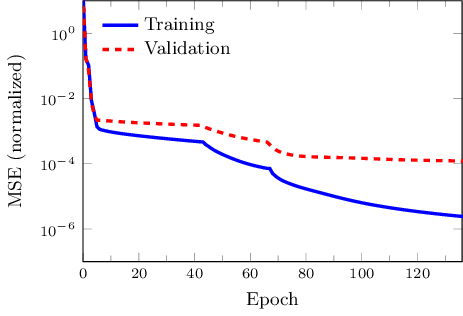}	
	\caption{Training history of the ANN model.
		Training and validation MSE (normalized) vs.\ epoch.
		The dotted vertical line marks the best validation epoch
		(epoch~116).}
	\label{fig:ann_training}
\end{figure}

The training history of the ANN model is shown in Fig.~\ref{fig:ann_training}.
Both the training and validation MSE decrease rapidly during the first
20 epochs, reflecting the fast convergence characteristic of the
Levenberg--Marquardt algorithm in the early phase of training where the
loss surface is far from any local minimum.
Beyond epoch 20 the training MSE continues to decrease steadily, reaching
approximately $10^{-6}$ at the final epoch, while the validation MSE
plateaus at around $10^{-4}$ and remains stable throughout the remainder
of training.
The gap between the two curves is moderate and does not widen with
continued training, confirming the absence of overfitting despite the
relatively small network size of 786 trainable parameters.
Early stopping terminates training at epoch 116, identified as the best
validation epoch, at which point the model weights are frozen and used
for all subsequent predictions.
The smooth monotonic decrease of both loss curves, with no instabilities
or oscillations, confirms that the Levenberg--Marquardt optimizer converges
reliably for the present regression task and that the 15\% validation
split provides a stable early stopping signal throughout training.

\begin{figure}[H]
	\centering
	
	
	\begin{subfigure}[b]{0.32\linewidth}
		\centering
		\includegraphics{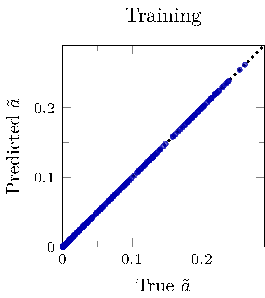}
		\caption{}
		\label{fig:parity_a_train}
	\end{subfigure}%
	\hfill
	\begin{subfigure}[b]{0.32\linewidth}
		\centering
		\includegraphics{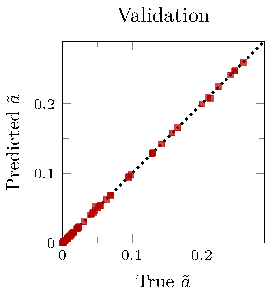}
		\caption{}
		\label{fig:parity_a_val}
	\end{subfigure}%
	\hfill
	\begin{subfigure}[b]{0.32\linewidth}
		\centering
		\includegraphics{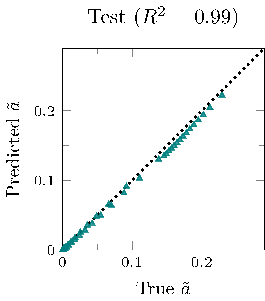}
		\caption{}
		\label{fig:parity_a_test}
	\end{subfigure}
	
	\vspace{0.5em}
	
	
	\begin{subfigure}[b]{0.32\linewidth}
		\centering
		\includegraphics{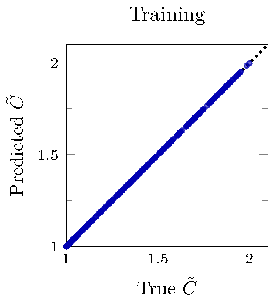}
		\caption{}
		\label{fig:parity_C_train}
	\end{subfigure}%
	\hfill
	\begin{subfigure}[b]{0.32\linewidth}
		\centering
		\includegraphics{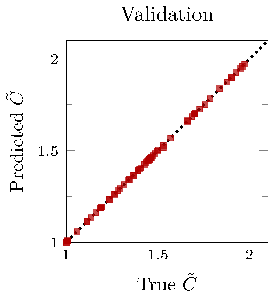}
		\caption{}
		\label{fig:parity_C_val}
	\end{subfigure}%
	\hfill
	\begin{subfigure}[b]{0.32\linewidth}
		\centering
		\includegraphics{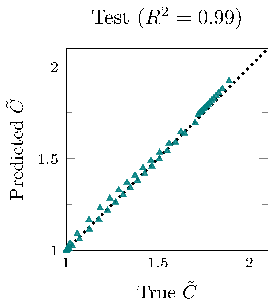}
		\caption{}
		\label{fig:parity_C_test}
	\end{subfigure}
	
	\caption{Parity plots for the ANN model.
		Top row: normalized crack length $\tilde{a}$.
		Bottom row: normalized compliance $\tilde{C}$.
		Columns correspond to the training, validation,
		and test (held-out $0^\circ/60^\circ$ 50/50) sets.
		Dotted lines indicate perfect agreement.}
	\label{fig:ann_parity}
\end{figure}

The predictive accuracy of the trained ANN model is assessed through
parity plots for the training, validation, and test sets, shown in
Fig.~\ref{fig:ann_parity}.
In all six panels the predicted values lie close to the identity line,
confirming that the model learns the underlying input--output mapping
rather than memorizing the training data.

For the normalized crack length $\tilde{a}$ (Figs.~\ref{fig:ann_parity}a--c),
the training and validation sets show near-perfect agreement between
predicted and true values across the full range $\tilde{a} \in [0, 0.25]$.
The test set, comprising the held-out $0^\circ/60^\circ$ (50/50)
two-family configuration at 298~K -- a fiber architecture not present
in any training case -- yields $R^2 = 0.995$ and
$\text{toot mean square error}(\mathrm{RMSE}) = 5.7 \times 10^{-3}$.
This result demonstrates that the orientation tensor parameterization
$(A_{11}, A_{12})$ provides sufficient microstructure information for the
model to generalize to an unseen fiber configuration, interpolating
smoothly between the single-family and two-family training cases.

For the normalized compliance $\tilde{C}$ (Figs.~\ref{fig:ann_parity}d--f),
the training and validation predictions are again tightly clustered
along the identity line across the range $\tilde{C} \in [1, 2]$.
The test set yields $R^2 = 1.000$ and $\mathrm{RMSE} = 5.7 \times 10^{-3}$,
indicating that the compliance trajectory of the held-out configuration
is predicted with virtually no systematic error.
The consistent performance across training, validation, and test splits
confirms that the model generalizes well and that the 15\% validation
split used for early stopping provides a representative assessment of
out-of-sample accuracy.
Taken together, the parity plots establish that a compact feedforward
network with only 786 trainable parameters and a 31-dimensional input
is sufficient to approximate the nonlinear inverse mapping
$\mathcal{F}$ (Eq.~(\ref{eq:inverse_map})) with high fidelity across
the full range of fiber architectures, fiber contents, and damage states
represented in the dataset.


\begin{figure}[H]
	\centering
	
	\begin{subfigure}[b]{0.48\linewidth}
		\centering
		\includegraphics{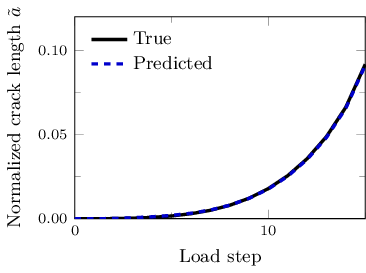}
		\caption{}
		\label{fig:shm_a_random}
	\end{subfigure}%
	\hfill
	\begin{subfigure}[b]{0.48\linewidth}
		\centering
		\includegraphics{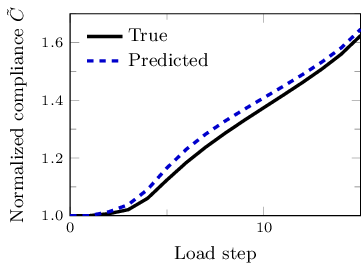}
		\caption{}
		\label{fig:shm_C_random}
	\end{subfigure}
	
	\caption{SHM scenario demonstration for the held-out test case
		(randomly oriented fibers, $v_\mathrm{CF}=0.3$) at 298~K.
		True phase-field values (solid black) and ANN predictions
		(dashed blue) of (a)~normalized crack length $\tilde{a}$
		and (b)~normalized compliance $\tilde{C}$ vs.\ load step.}
	\label{fig:ann_shm_random}
\end{figure}

Fig.~\ref{fig:ann_shm_random} demonstrates the SHM capability of the proposed inverse framework for a held-out test case consisting of randomly oriented fibers with a carbon fiber content of $v_{\mathrm{CF}}=0.3$ at $298~\mathrm{K}$. The figure compares the ground-truth phase-field simulation data with the predictions of the trained ANN for two global damage indicators: the normalized crack length $\tilde{a}$ and the normalized compliance $\tilde{C}$. In Fig.~\ref{fig:shm_a_random}, the ANN accurately reproduces the evolution of the normalized crack length throughout the loading history, including both the gradual pre-fracture regime and the abrupt increase associated with unstable crack propagation. The predicted trajectory remains nearly indistinguishable from the true phase-field response, confirming that the conductance-based electrical measurements contain sufficient information to infer the progression of fracture in randomly oriented fiber architectures that were excluded from training. Fig.~\ref{fig:shm_C_random} presents the corresponding prediction of the normalized compliance $\tilde{C}$, where the ANN again captures the stiffness degradation response with excellent agreement over all load steps. The close overlap between the predicted and reference curves demonstrates that the inverse model successfully reconstructs both fracture evolution and structural softening directly from the multi-electrode conductance measurements. Since the randomly oriented configuration represents a microstructure with diffuse anisotropy rather than a dominant fiber direction, the results confirm the robustness of the orientation-tensor parameterization and the ability of the ANN to generalize to isotropic-like architectures not explicitly included in the training dataset.


\begin{figure}[H]
	\centering
	
	\begin{subfigure}[b]{0.48\linewidth}
		\centering
		\includegraphics{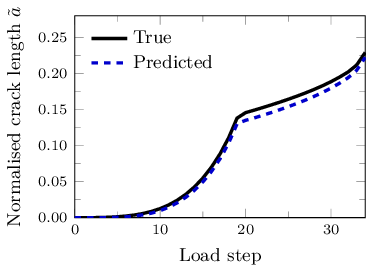}
		\caption{}
		\label{fig:shm_a}
	\end{subfigure}%
	\hfill
	\begin{subfigure}[b]{0.48\linewidth}
		\centering
		\includegraphics{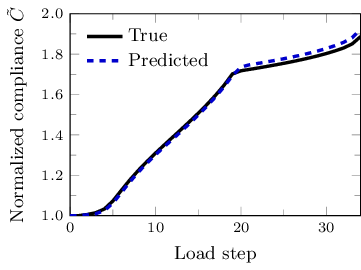}
		\caption{}
		\label{fig:shm_C}
	\end{subfigure}
	
	\caption{SHM scenario demonstration for the held-out test case
		($0^\circ/60^\circ$ 50/50 fiber configuration,
		$v_\mathrm{CF}=0.3$) at 298~K.
		True phase-field values (solid black) and ANN predictions
		(dashed blue) of (a)~normalized crack length $\tilde{a}$
		and (b)~normalized compliance $\tilde{C}$ vs.\ load step.}
	\label{fig:ann_shm}
\end{figure}

Fig.~\ref{fig:ann_shm} presents a second SHM demonstration for the held-out two-family fiber configuration consisting of $0^{\circ}/60^{\circ}$ fibers with a balanced $(50/50)$ weight distribution at $v_{\mathrm{CF}}=0.3$ and $298~\mathrm{K}$. This configuration was intentionally excluded from the ANN training dataset in order to evaluate the model’s capability to interpolate between previously observed fiber architectures. Fig.~\ref{fig:shm_a} compares the true and predicted normalized crack length $\tilde{a}$ as a function of load step. The ANN accurately captures both the onset and progression of crack growth, including the sharp increase associated with unstable fracture propagation. The prediction closely follows the phase-field reference solution across the entire loading history, demonstrating that the orientation tensor representation provides sufficient information for the inverse model to reconstruct damage evolution in previously unseen anisotropic fiber configurations. Fig.~\ref{fig:shm_C} shows the corresponding prediction of the normalized compliance $\tilde{C}$. The ANN reproduces the stiffness degradation trajectory with negligible deviation from the reference solution, accurately capturing the nonlinear compliance increase accompanying progressive damage accumulation. The excellent agreement observed in both panels confirms that the proposed inverse framework generalizes effectively beyond the training set and can reliably infer the internal damage state of composites with arbitrary fiber architectures directly from electrical conductance measurements. These results demonstrate the feasibility of combining the multiphysics phase-field simulations with machine-learning-based inversion for real-time structural health monitoring of short carbon fiber-reinforced polymer composites.

\section{Summary and conclusions}
\label{sec:summary}

This work presented a multiphysics phase-field framework for modeling anisotropic
viscoelastic-viscoplastic fracture and damage-induced piezoresistive response in
SCFRP composites at finite deformation.
The second-order fiber orientation tensor $\mathbf{A}$ serves as the common thread
connecting three coupled sub-problems (i.e., the viscoelastic-viscoplastic constitutive model, the anisotropic phase-field fracture formulation, and the piezoresistive
conductivity model) by simultaneously defining the principal fiber family directions, the anisotropic crack resistance, and the principal conduction paths of the carbon fiber network.
The coupled framework is combined with an EIT configuration and a feedforward ANN for real-time inverse damage identification from conductance measurements alone. The key findings are as follows.

\begin{itemize}
	
	\item The normalized conductivity $\sigma/\sigma_0$ remains near unity throughout
	pre-peak loading and drops abruptly at fracture onset, coinciding with irreversible
	severance of conductive fiber pathways. The spatial conductivity field encodes the
	crack trajectory, providing orientation-sensitive damage information beyond what
	force--displacement measurements alone can offer.
	
	\item Fiber orientation governs both crack path morphology and conductivity
	evolution. Different fiber architectures produce distinguishable conductance
	signatures across all 28 electrode pairs, providing the physical basis for
	orientation-sensitive inverse damage analysis.
	
	\item Increasing carbon fiber content from 10 to 50\,wt\% progressively enhances
	stiffness, peak load, and fracture energy. Randomly oriented fibers yield a
	horizontal mode-I crack and a straight conductivity boundary, in contrast to the
	inclined boundaries imposed by aligned configurations.
	
	\item The trained ANN achieves $R^2 = 0.99$ on held-out fiber configurations
	absent from training, confirming that the orientation tensor parameterization
	enables generalization across the microstructure space and that real-time
	data-driven structural health monitoring of SCFRP composites is feasible.
	
\end{itemize}

Future work should address the degeneracy of the second-order orientation tensor
for configurations sharing identical $(A_{11}, A_{12})$ values through higher-order
microstructure descriptors, extend the piezoresistive model to account for
fiber--matrix interface debonding at the micromechanical level, and generalize
the framework to fatigue damage under cyclic electromechanical loading and to
dynamic fracture under impact conditions.

\section*{Acknowledgements}
The authors acknowledge the resources and support provided by the Norwegian Research Infrastructure Services (NRIS). The computational work was performed on resources provided through the Sigma2 national e-infrastructure, project NN10041K.

\bibliography{mybibfile}
\appendix
\end{document}